\documentclass[iop]{emulateapj}

\def\msun{\hbox{M$_{\odot}$}}
\def\mdot{\hbox{$\dot M$}}
\def\micron{$\mu$m}
\def\microns{$\mu$m}

\newcommand\be{\begin{equation}}
\newcommand\en{\end{equation}}

\begin{document}

\shortauthors{Espaillat et al.}
\shorttitle{Pre-Transitional Disks}

\title{Unveiling the Structure of Pre-Transitional Disks}

\author{C. Espaillat\altaffilmark{1,2}, 
P. D'Alessio\altaffilmark{3},
J. Hern\'{a}ndez\altaffilmark{4}, 
E. Nagel\altaffilmark{5}, 
K. L. Luhman\altaffilmark{6}, 
D. M. Watson\altaffilmark{7},
N. Calvet\altaffilmark{8},  
J. Muzerolle\altaffilmark{9}, 
\& M. McClure\altaffilmark{8}}

\altaffiltext{1}{Harvard-Smithsonian Center for Astrophysics, 60 Garden
Street, MS-78, Cambridge, MA, 02138, USA; cespaillat@cfa.harvard.edu}
\altaffiltext{2}{NSF Astronomy \& Astrophysics Postdoctoral Fellow}
\altaffiltext{3}{Centro de Radioastronom\'{i}a y Astrof\'{i}sica,
Universidad Nacional Aut\'{o}noma de M\'{e}xico, 58089 Morelia,
Michoac\'{a}n, M\'{e}xico; p.dalessio@crya.unam.mx}
\altaffiltext{4}{Centro de Investigaciones de Astronom\'{i}a (CIDA),
Merida, 5101-A, Venezuela; jesush@cida.ve} 
\altaffiltext{5}{Departamento
de Astronom\'\i a, Universidad de Guanajuato, Guanajuato, Gto, M\'exico
36240; erick@astro.ugto.mx} 
\altaffiltext{6}{Department of Astronomy and
Astrophysics, The Pennsylvania State University, University Park, PA
16802, USA; kluhman@astro.psu.edu} 
\altaffiltext{7}{Department of Physics and
Astronomy, University of Rochester, NY 14627-0171, USA;
dmw@pas.rochester.edu} 
\altaffiltext{8}{Department of
Astronomy, University of Michigan, 830 Dennison Building, 500 Church
Street, Ann Arbor, MI 48109, USA; ncalvet@umich.edu, melisma@umich.edu}
\altaffiltext{9}{Space Telescope Institute, 3700 San Martin Drive,
Baltimore, MD 21218, USA; muzerol@stsci.edu}

\begin{abstract}

In the past few years, several disks with inner holes that are
relatively empty of small dust grains have been detected and are known
as transitional disks.  Recently, {\it Spitzer} has identified a new
class of ``pre-transitional disks" with gaps based on near-infrared
photometry and mid-infrared spectra; these objects have an optically
thick inner disk separated from an optically thick outer disk by an
optically thin disk gap. A near-infrared spectrum provided the first
confirmation of a gap in the pre-transitional disk of LkCa~15 by
verifying that the near-infrared excess emission in this object was due
to an optically thick inner disk. Here we investigate the difference
between the nature of the inner regions of transitional and
pre-transitional disks using the same veiling-based technique to extract
the near-infrared excess emission above the stellar photosphere.
However, in this work we use detailed disk models to fit the excess
continua as opposed to the simple blackbody fits used previously. We
show that the near-infrared excess emission of the previously identified
pre-transitional disks of LkCa~15 and UX~Tau~A in the Taurus cloud as
well as the newly identified pre-transitional disk of Rox~44 in
Ophiuchus can be fit with an inner disk wall located at the dust
destruction radius. We also present detailed modeling of the broad-band
spectral energy distributions of these objects, taking into account the
effect of shadowing by the inner disk on the outer disk,
but considering the finite size of the star, unlike other recent treatments. 
The
near-infrared excess continua of these three pre-transitional disks,
which can be explained by optically thick inner disks, are significantly
different from that of the transitional disks of GM~Aur, whose
near-infrared excess continuum can be reproduced by emission from
sub-micron-sized optically thin dust, and DM~Tau, whose near-infrared
spectrum is consistent with a disk hole that is relatively free of small
dust. The structure of pre-transitional disks may be a sign of young
planets forming in these disks and future studies of pre-transitional
disks will provide constraints to aid in theoretical modeling of planet
formation.

\end{abstract}

\keywords{accretion disks, stars: circumstellar matter, planetary
systems: protoplanetary disks, stars: formation, stars: pre-main
sequence}

\section{Introduction} \label{intro}

Several disks which have nearly photospheric near-infrared emission but
substantial excesses above the stellar photosphere at wavelengths beyond
$\sim$20~{\micron} have been observed and are referred to as
``transitional disks'' \citep{strom89}. Using data from the {\it
Spitzer} Infrared Spectrograph \citep[IRS;][]{houck04}, detailed
modeling has demonstrated that this flux deficit at near-infrared
wavelengths relative to full disks can be explained by optically thick
disks with inner holes of $< 40$~AU. In most cases these inner holes are
not completely devoid of material (e.g. GM~Aur, TW Hya, CS Cha, CVSO
224); a minute amount of micron- or sub-micron-sized optically thin dust
exists within the hole, producing a small infrared excess over the
photospheric flux, still well below the median excess of Class II
objects, as well as silicate emission \citep{calvet02,calvet05,
espaillat07a, espaillat08b}.  Gas has also been detected within the
inner holes of transitional disks
\citep[e.g.][]{najita03,bergin04,salyk07}.

Recently, the {\it Spitzer Space Telescope} \citep{werner04} identified
a new class of disks called ``pre-transitional disks" around LkCa~15 and
UX~Tau~A \citep{espaillat07b}. These disks have deficits of mid-infrared
flux (5--20~{\micron}) and substantial excesses at longer wavelengths,
as is seen in the transitional disks. However, in contrast to the small
or absent near-infrared (2--5~{\micron}) excesses exhibited by
transitional disks, pre-transitional disks have significant
near-infrared excesses relative to their stellar photospheres, similar
to the median spectral energy distribution (SED) of disks in Taurus
\citep{dalessio99}.  The distinctive shapes of these SEDs indicate that
pre-transitional disks have an inner disk separated from an outer disk
and that we may be seeing the development of gaps within protoplanetary
disks.

While the truncation of LkCa~15's outer disk has been imaged in the
millimeter \citep{pietu06}, \citet{espaillat07b} showed that the
substantial near-infrared excess of LkCa~15 could be explained by either
optically thick material or by ${\sim}10^{-11}$ M$_{\odot}$ of optically
thin dust mixed with the gas in the inner disk.  In order to resolve
this issue, \citet{espaillat08a} obtained a medium resolution
near-infrared spectrum spanning the wavelength range 2--5~{\micron}.
This near-infrared spectrum had absorption lines that were weaker
relative to the spectrum of a standard star of the same spectral type.
This phenomenon, known as ``veiling" \citep{hartigan89}, is also
observed in similar spectra of full disks and is due to emission from
dust located at the dust sublimation radius \citep{muzerolle03}.
\citet{espaillat08a} measured a veiling factor (r$_{K}$) of 0.3$\pm$0.2
for LkCa~15 at $\sim$2.2~{\micron} and fit the near-infrared excess from
2--5~{\micron} with a single-temperature blackbody of 1600~K.  This
behavior can be explained by an optically thick wall located at the dust
sublimation radius, as is seen in full disks \citep{muzerolle03}. These
data confirmed that LkCa~15 has an inner optically thick disk, making
this observation the first independent verification of a gap in a
protoplanetary disk.

Here we expand our sample to include the pre-transitional disks of UX Tau A
\citep{espaillat07b} and Rox~44 \citep{furlan09},  and the transitional
disks of GM~Aur and DM~Tau \citep{calvet05}  in order to explore the
structure of the inner regions of pre-transitional and transitional
disks.  To do this we use veiling measurements in the K-band to extract
the near-infrared (-IR) excess emission of these objects and then fit
this emission with disk models \citep{dalessio05,calvet02}. Given the
sensitivity of veiling measurements on the adopted spectral type, we
redetermined spectral types for our targets using optical spectra
({\S}~\ref{sec:spt}). We also tested our veiling measurement methods
with the spectrum of the diskless, weak-line T Tauri star (WTTS) LkCa~14
(\S~\ref{sec:lkca14}).

We find that the near-infrared spectra of the pre-transitional disks of
LkCa~15, UX~Tau~A, and Rox~44 are well-explained by the wall of an
optically thick inner disk ({\S}~\ref{sec:lkca14}, {\S}~\ref{sec:ux},
{\S}~\ref{sec:rox}). In contrast, our data shows that the inner hole of
the transitional disk of GM~Aur contains a small amount of optically
thin sub-micron-sized dust while DM~Tau's hole is relatively free of
small dust ({\S}~\ref{sec:gm} \& {\S}~\ref{sec:dm}). Our results are
consistent with veiling and interferometric measurements found in the
literature. We also perform detailed model fits to the broad-band SEDs
of our pre-transitional disk sample, and explore the effect of shadowing
of the outer disk by the inner disk ({\S}~\ref{sec:models}), 
taking into account that the star is not a point
source, as has been adopted in other studies \citep{espaillat07b, mulders10}.
The
structure of pre-transitional disks suggests that of the disk clearing
mechanisms proposed to date, planet formation
\citep[e.g.][]{goldreich80, rice03, varniere06} is most likely a
dominant factor in clearing these disks.

\section{Observations \& Data Reduction} \label{redux}

Near-infrared spectra of LkCa~14, LkCa~15, UX~Tau~A, Rox~44, GM~Aur, and
DM~Tau  were obtained at the NASA Infrared Telescope Facility (IRTF)
facility using SpeX \citep{rayner03}.  For each of our targets, we used
the Long XD2.1 grating, covering 2.1 to 5.0~{\micron}, with a slit of
$0\farcs5$$\times$15$^\prime$$^\prime$ for a resolution
($\lambda$/$\delta$$\lambda$) of 1500. To trace near-IR emission at
shorter wavelengths, if present, we obtained additional spectra from
$\sim$0.8--2.5~{\micron} for LkCa~15, UX~Tau~A, Rox~44, and GM~Aur using
the low-resolution prism with a slit of
$0\farcs8$$\times$15$^\prime$$^\prime$ ($\lambda$/$\delta$$\lambda$$\sim$250)
 with an exposure time of 40
seconds. The dates of the observations, the LXD exposure times, and the
signal-to-noise ratios for each target are given in
Table~\ref{tab:spexlog}.  We note that the prism data for Rox~44 were
first presented in \citet{mcclure10}.

We extracted the data with Spextool \citep{cushing04} and corrected for
telluric absorption using the {\it xtellcor} routine \citep{vacca03}.
Our telluric standards for LkCa~14, LkCa~15, DM~Tau, GM~Aur, UX~Tau~A,
and Rox~44 were HD25175, HD27777, HD27761, HD27777, HD27777, and
HD146606 respectively. Bad pixels and regions of high telluric noise
(e.g. 2.5--2.9~{\micron} and 4.2--4.6~{\micron}) were manually removed
from the spectra.

To measure the spectral types for objects in our sample, we used
archival low-dispersion optical spectra obtained at the 1.5~m telescope
of the Whipple Observatory with the Fast Spectrograph for the
Tillinghast Telescope \citep[FAST; ][]{fabricant98} equipped
with the Loral 512$\times$2688 CCD. The spectrograph was set up in the
standard configuration used for FAST COMBO projects: a 300 groove
mm$^{-1}$ grating and a 3{\arcsec} wide slit. This combination offers
spectral coverage over $\sim$3600--7500~{\AA}, with a resolution of
$\sim$6~{\AA}. All spectra were reduced at the Harvard-Smithsonian
Center for Astrophysics using software developed specifically for FAST
COMBO observations and were wavelength-calibrated and combined using
standard IRAF routines. Data for UX~Tau~A and LkCa~14  were obtained in
1995 and 1996 as part of Program 38 (PI: Brice{\~n}o) and are publicly
available in the FAST
database.\footnote{http://tdc-www.harvard.edu/cgi-bin/arc/fsearch}
Spectra for LkCa~15, DM~Tau, and GM~Aur were obtained in 1995 and 1996
as part of Program 30 (PI: Kenyon) and were first used in
\citet{kenyon98} to measure the equivalent widths of H$_{\alpha}$, [N
II], He I, and [S II] lines. We note that no FAST spectra were available
for Rox~44.

\section{Disk Model} \label{sec:mod}

In the next section, we fit the near-IR excess emission and the
broad-band SEDs of some of the objects in our sample with disk models.
Below we describe the models used to reproduce the observed emission.

In the pre-transitional disks, there is an inner wall located at the
dust sublimation radius which dominates the near-IR (2--5~{\micron})
emission. Here we use the wall model of \citet{dalessio05}. The wall is
assumed to be vertical\footnote{We leave it to future work to explore
the curvature of the inner wall (Nagel et al. in prep).} with evenly
distributed dust. The wall's optical depth increases radially and is
optically thin closest to the star. The stellar radiation impinges
directly onto the wall, with an angle of 0$^{\circ}$ to the normal of
the wall's surface. Following \citet{calvet91,calvet92}, the radial
distribution of temperature for the wall atmosphere is
\begin{equation} T_d(\tau_d)^4 \sim \frac{F_0}{4\sigma_R} \left [
2 + {\frac{\kappa_s} {\kappa_d}} e^{-q\tau_d} \right ]
\label{walltdistrib}
\end{equation} 
where $F_0=(L_{*}+L_{acc})/4\pi R_{wall}^{2}$, $\sigma_R$ is the
Stefan-Boltzmann constant, $\tau_d$ is the total mean optical depth in
the disk, and $\kappa_s$ and $\kappa_d$ are the mean opacities to the
incident and local radiation, respectively. $L_* $ is the stellar
luminosity and $L_{acc}$ ($\sim GM_*\mdot/R_*$) is the luminosity of the
accretion shock onto the stellar surface, which also heats the inner
wall. To derive the location of the wall, one sets $\tau_d = 0$ and
\begin{equation}\label{walleqn} R_{wall} \sim 
\left [{\frac{(L_* + L_{acc})}{16 \pi \sigma_R} }
( 2 + { \frac{\kappa_s} {\kappa_d} }) \right ] ^ {1/2} 
{1 \over T_{wall}^2 }.
\label{rdust}
\end{equation} 
$T_{wall}$ is the temperature at the surface of the optically thin wall
atmosphere (i.e. T$_d$($\tau_d$=0)). When fitting the wall located at
the dust sublimation radius, the relevant dust grains located at these
high temperatures (1000--2000~K) are composed of silicates. Following
\citet{dalessio05}, here we adopt silicates with a dust-to-gas mass
ratio (${\zeta}_{sil}$) of 0.0034 using opacities from
\citet{dorschner95}. The grain size distribution used here (and
throughout the disk) follows the form $a^{-3.5}$ where $a$ varies
between $a_{min}$ and $a_{max}$ \citep{mathis77}. The temperature of the
inner wall (T$_{wall}^i$), the height of the wall (z$_{wall}^i$), and
the maximum grain size are adjusted to fit the SED.

In the transitional disks, there is an inner hole that sometimes
contains a small amount of optically thin dust which contributes to the
SED between $\sim$2--10~{\micron}. We calculate the emission from this
optically thin dust region by integrating the flux from optically thin
annuli where all of the dust grains are heated by stellar radiation
following \citet{calvet02}. Some pre-transitional disks also contain
optically thin dust within their disk gaps.

In both pre-transitional and transitional disks, there is an outer wall
located where the outer disk is truncated and this wall dominates the
SED emission from $\sim$20--30~{\micron}. The temperature structure of
this wall is calculated in the same manner as the structure of the inner
wall above. For the wall of the outer disk, we add organics and troilite
to the dust mixture following \citet{dalessio05} with ${\zeta}_{org}$ =
0.001 and ${\zeta}_{troi}$ = 0.000768 and sublimation temperatures of
$T_{org}$ = 425~K and $T_{troi}$ = 680~K. We include water ice as well
where the local temperature is below the sublimation temperature of this
component (T$_{ice}$= 110~K).  We use ${\zeta}_{ice}$ = 0.00056, which
is 10 $\%$ of the abundance proposed by \citet{pollack94}, since
\citet{dalessio06} found that \citet{pollack94}'s ice abundance produces
features that are not observed in typical disk SEDs. Opacities for
organics, troilite, and water ice are adopted from \citet{pollack94},
\citet{begemann94}, and \citet{warren84}. In this paper we do not study the detailed composition of
the dust, but simply illustrate that a typical dust composition can
reasonably explain the observed SED. We adopt an ISM-sized grain
distribution \citep[i.e. $a_{min}$=0.005~{\micron} and
$a_{max}$=0.25{\micron};][]{mathis77} and varied the outer wall
temperature (T$_{wall}^o$) and height (z$_{wall}^o$) to achieve the
best-fit to the SED.

In the pre-transitional disks, the inner optically thick disk will cast
a shadow on the outer disk.  Assuming that the central star is a point
source, this shadowing of the outer disk would be substantial. 
On the contrary,
the star appears as a finite source to the outer disk and so the inner
disk will cast a large penumbra and a small umbra on the outer disk.  If
a point on the outer wall is in the umbra, it does not see the star. If
it is in the penumbra, it will see some of the star, and will still be
illuminated to some extent. If the point on the wall is completely out
of the shadow (both umbra and penumbra), it is fully illuminated by the
star. In calculating the outer wall's emission we take into account
shadowing by a finite source star (see Appendix).
Since we assume the wall is vertical and the luminosity along the the
outer wall changes due to the geometry of the shadow, there will be a
range of temperatures along the surface of the wall, consistent with the
varying degree of illumination.  We calculate the resulting SED with
this range of temperatures and the radius of the wall is determined by
the temperature in the fully illuminated part.

Both pre-transitional and transitional disks have a contribution to
their SEDs beyond $\sim$40~{\micron} from the outer disk  and we use the
irradiated accretion disk models of \citet{dalessio06} to model the
emission of the disk behind the wall. To simulate grain growth and
settling, the disk is composed of two dust grain size distributions. In
the upper disk layers, grains are ISM-sized and in the disk midplane the
maximum grain size is 1~mm \citep{dalessio06}. The viscosity parameter
($\alpha$) and settling parameter ($\epsilon$; i.e. the dust-to-gas mass
ratio in the upper disk layers relative to the standard dust-to-gas mass
ratio) are varied to achieve the best-fit to the SED. We adopt an outer
disk radius of 300~AU for all of our disks.

\section{Analysis} \label{sec:ana}

 Classical T Tauri stars tend to have observed photospheric absorption
 lines which are weaker than those of non-accreting stars.  This
 ``veiling''  is due to an excess continuum which adds to the star's
 emission and ``fills in'' stellar absorption lines, making them appear
 weaker when compared to a standard star of the same spectral type
 \citep{hartigan89}. In the optical and shorter wavelengths, this is
 primarily due to the hot continuum which arises from the accretion
 shocks onto the star \citep{calvet98}.  In the near-infrared, for stars
 which are not strong accretors, veiling of photospheric absorption
 lines is mainly due to dust in the disk \citep{muzerolle03}.

 Since veiling measurements are dependent on the spectral types that are
 adopted, we first re-derive spectral types for most of our sample using FAST optical
 spectra  in {\S}~\ref{sec:spt}. 
Then, in 
{\S}~\ref{sec:nearir} we measure near infrared veilings for our sample,
extract the excess emission, and fit it with disk wall models.
We
 then model the broad-band SEDs of the pre-transitional disks of
 LkCa~15, UX~Tau~A, and Rox~44 with the best-fit inner wall model,
 including the shadowing of this inner wall on the outer disk.

\subsection{Spectral Type Classification} \label{sec:spt}
  
Spectral types were derived with the SPTCLASS
tool,\footnote{http://www.astro.lsa.umich.edu/$\sim$hernandj/SPTclass/\\
sptclass.html} 
an IRAF$/$IDL code based on the methods described by
\citet{hernandez04}. The original code was optimized to calculate the
spectral types of early type stars (B, A, \& F)  based on 33 spectral
features that are sensitive to changes in ${\rm T_{eff}}$. SPTCLASS was
later revised to incorporate 11 spectral indices optimized for solar
type stars (from middle F to early K) and 16 spectral indices optimized
for low mass stars  (from early K to M5). 
Each spectral index is obtained by measuring the decrease in flux
from what would be expected at the line center by interpolating between
two adjacent bands on each side of the line. Each index is calibrated
using spectroscopic standards observed with FAST.

Most of the spectral classifications derived in this work
differ somewhat from those reported in the
literature. Our classification of DM~Tau (M1.5$\pm$1.0) is in agreement
with previous measurements of M0.5 \citep{cohen79} and M2
\citep{herbig77}. However, we measure a spectral type for LkCa~14 of
K5.5$\pm$1.0 while \citet{herbig86} derived a spectral type of M0. For
GM~Aur, we derive a spectral type of K5.5$\pm$1.0, but in the literature
we find spectral classifications of K3 \citep{herbig77} and K7--M0
\citep{cohen79} for this object. 
We derive a spectral type of K3$\pm$1.0 for
LkCa 15 which differs from the K5 spectral type for this object reported by \citet{herbig86}.
We classify UX Tau~A as a G8.0$\pm$2.0
star; \citet{herbig77} and \citet{cohen79} both reported a spectral
type of K2,  \citet{rydgren76} reported G5, and \citet{hartigan94} find
K5--K2. Since no FAST spectra are available for Rox~44, we adopt a
spectral type of K3 from \citet{bouvier92} for this object.

The spectral types derived here (Table~\ref{tab:spexchar}) are based on
a greater number of spectral indices than used in the works cited above.
Since commonly used spectral indices such as He I (4922~{\AA},
5016~{\AA}) and Na I (5890~{\AA}, 5896~{\AA}) can be affected by
emission or anomalous absorption, it is crucial to base spectral type
measurements on several indices \citep[see discussion
in][]{hernandez04}. \citet{herbig77}, \citet{herbig86},
\citet{rydgren76}, and \citet{cohen79} analyze spectral ranges of
5850--6700~{\AA}, 5800--6800~{\AA}, 3500--7000~{\AA}, and
4270--6710~{\AA}, respectively, with a resolution similar to that of
our observations, but with fewer spectral indices. The medium resolution
($\sim$2~{\AA}) spectra by \citet{hartigan94} cover a much smaller range
in wavelength (5700-6900~{\AA} and 5860-6890~{\AA}) that includes few
spectral indices.  We include more indices and these are 
based on a library of standards obtained with the same
instrument and configuration.

\begin{figure}
\epsscale{1}
\plotone{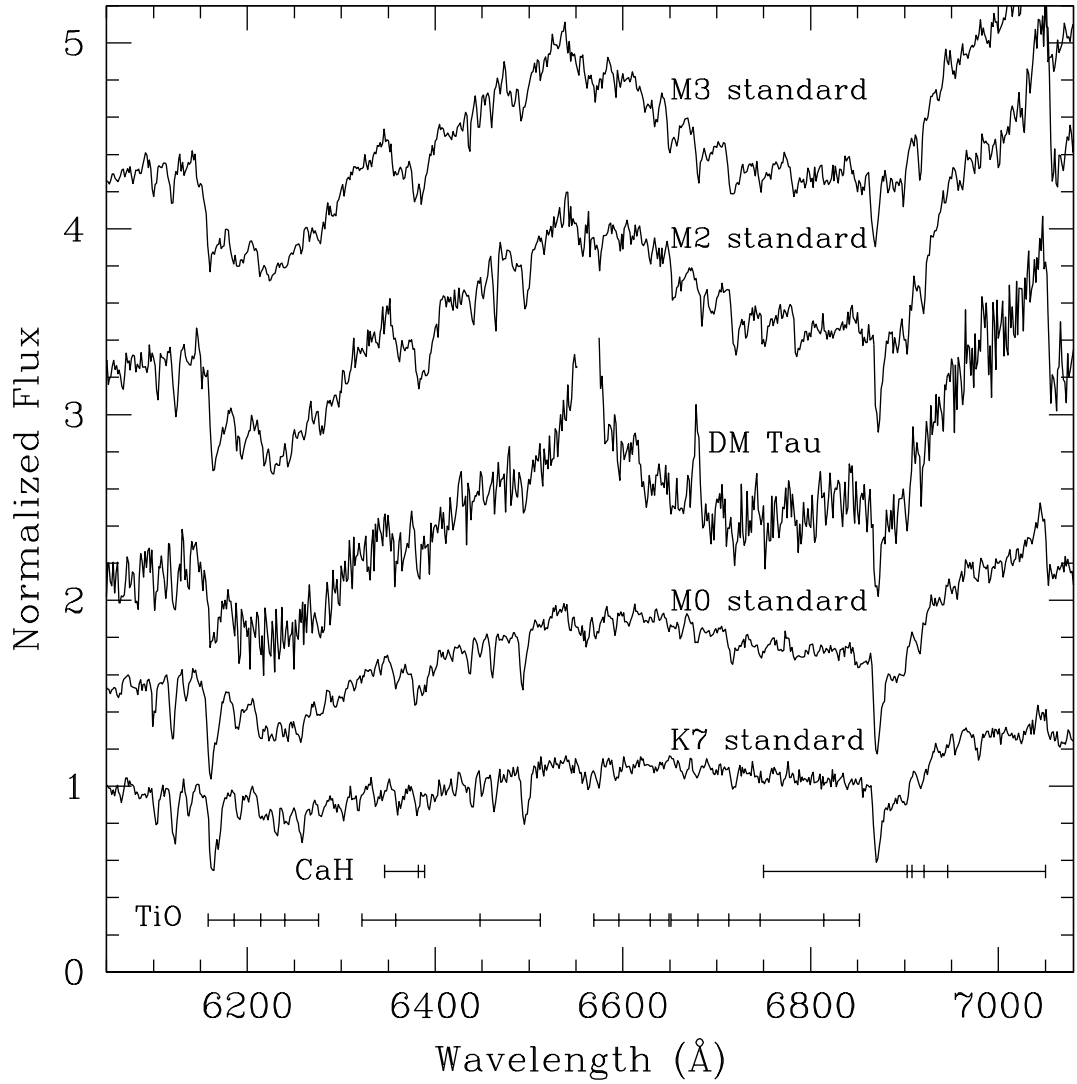}
\caption[FAST spectrum of DM~Tau compared to spectral type
standards]{FAST spectrum of DM~Tau compared to spectral type standards.
We measure a spectral type of M1.5$\pm$1 for DM~Tau. Some of the
features used for spectral classification are marked.  The strong
H$_{\alpha}$ emission feature at $\sim$6550~{\AA} is cut off for clarity.
}
\label{figsptdm}
\end{figure}

\begin{figure}
\epsscale{1}
\plotone{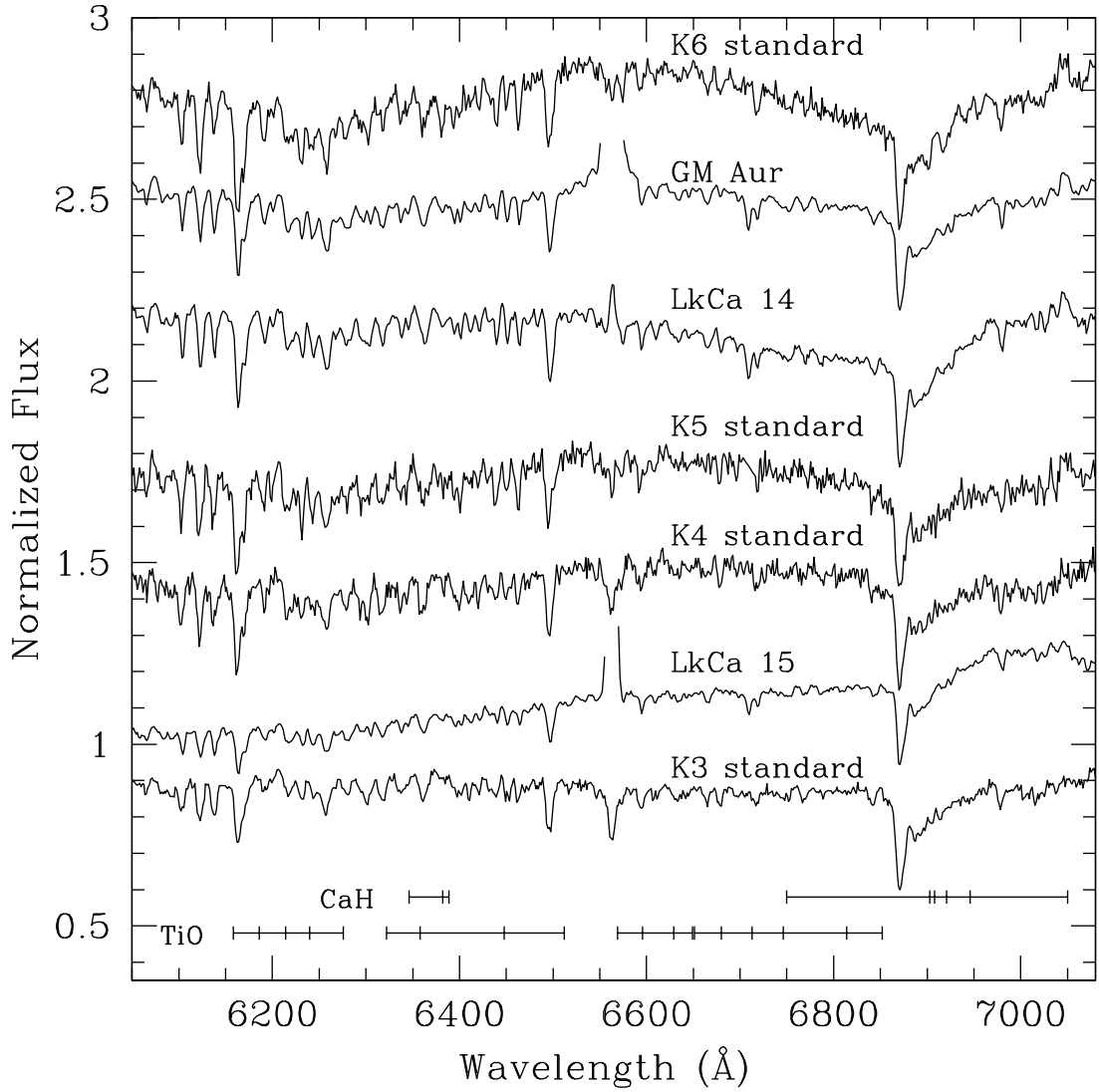}
\caption[FAST spectra of GM~Aur, LkCa~14, and LkCa~15 compared to
spectral type standards]{FAST spectra of GM~Aur, LkCa~14, and LkCa~15
compared to spectral type standards. We measure a spectral type of
K5.5$\pm$1 for both GM~Aur and LkCa~14 and a spectral type of K3$\pm$1
for LkCa~15. Some of the features used for spectral classification are
marked. The strong H$_{\alpha}$ emission features at $\sim$6550~{\AA} in GM~Aur and
LkCa~15 have been trimmed.
}
\label{figsptgm}
\end{figure}

\begin{figure}
\epsscale{1}
\plotone{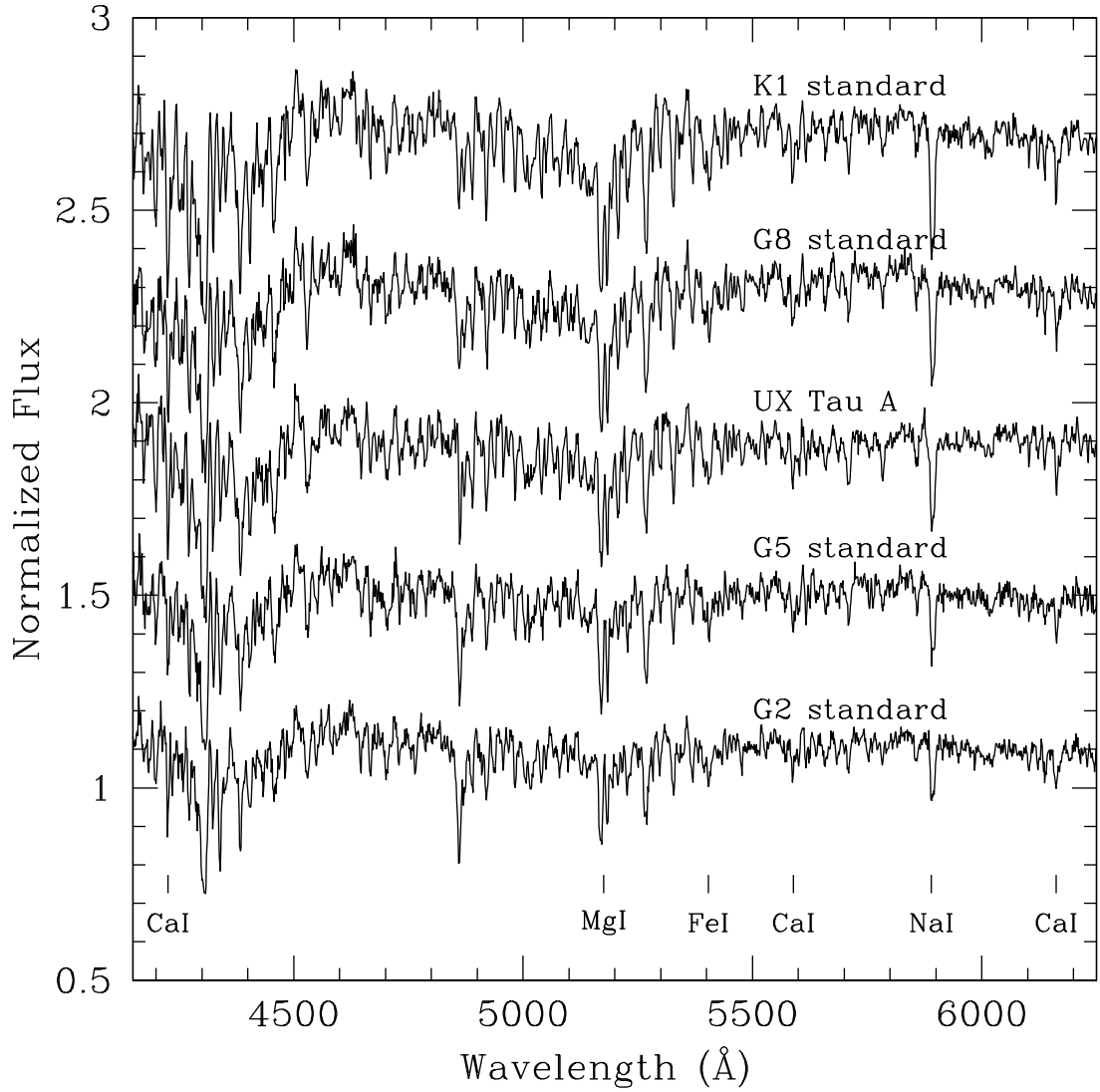}
\caption[FAST spectrum of UX~Tau~A compared to spectral type
standards]{FAST spectrum of UX~Tau~A compared to spectral type
standards. We measure a spectral type of G8$\pm$2 for UX~Tau~A. Some of
the features used for spectral classification are marked.
}
\label{figsptux}
\end{figure}

In Figures~\ref{figsptdm},~\ref{figsptgm}, and~\ref{figsptux} we
illustrate the accuracy of the SPTCLASS fitting procedure by comparing
our targets to spectral type standards. In the late-type stars, 
the molecular bands of TiO and CaH were used in classifying the
targets (Figures~\ref{figsptdm} and~\ref{figsptgm}). In G-type stars,
the metallic lines play an important role in spectral classification and
the features used to classify UX~Tau~A include the Ca I, Mg I, 
Fe I, and Ca I lines (Figure~\ref{figsptux}). The TiO and CaH bands in LkCa~14
(Figure~\ref{figsptgm}) are in better agreement with a K5$/$K6 star. In
GM~Aur and LkCa 15, these bands are similar to those in a K5$/$K6 star and
 K4$/$K3 star, respectively,
rather than earlier (Figure~\ref{figsptgm}) or later
(Figure~\ref{figsptdm}) type stars. The lines in UX~Tau~A, particularly
Ca I and Na I, are weaker than those in a later type star (e.g. K1) and
resemble the depths of the lines in the G8 star but not the earlier type
stars (Figure~\ref{figsptux}). Since there are no G6 or G7 objects in
the library of standards used for our classification we cannot exclude
it is a G6 or G7 star and so the uncertainty in UX~Tau~A's G8 spectral type
classification is $\pm$2.0.  We note that our SpeX spectra are consistent
with the spectral types derived with FAST.

\subsection{Near-Infrared Excess Emission} \label{sec:nearir} 

 To estimate the amount of veiling present in our sample, we compared
 the line strengths in the target to those of a standard star of the
 same spectral type ({\S}~\ref{sec:nearir}). 
 The spectra of corresponding template
 stars were obtained from the IRTF Spectral Library \citep{cushing05,
 rayner09}\footnote
 {http://irtfweb.ifa.hawaii.edu/$\sim$spex/IRTF$\_$Spectral$\_$Library/}. 
 We first normalized the
 target and standard stars' emission at 2.2~{\micron} and then
 artificially veiled the standard spectrum by adding a flat continuum of
 $F_{excess} = r_{K} \times F_{standard}$ where $r_{K}$ is the veiling
 factor (=$F_{excess}$$/$$F_{*}$). The standard spectrum was veiled
 until the depths of the absorption lines matched those of the target
 spectrum between 2.20 -- 2.28~{\micron}. We
 normalized the standard spectrum in the K-band to the 
colors of a corresponding stellar
photosphere from \citet{kh95} scaled to the observed J-band flux.
We then scaled the target by
 $1+ r_{K}$ and subtracted the standard spectrum to extract the shape of
 near-infrared excess emission.  When excess emission is present, we fit
 the continuum with a disk model and derive the best-fit parameters.

\subsubsection{LkCa~14} \label{sec:lkca14}

To test our method of measuring the veiling and excess emission of T
Tauri stars (TTS), we analyzed the spectrum of LkCa~14,  a single
\citep{white01}, non-accreting diskless T Tauri star in Taurus
\citep{kh95}.  The expectation is that this object should have no K-band
veiling since it does not exhibit excess emission at near- and mid-IR
wavelengths in its SED \citep{kh95}.  \citet{karr10} report a tentative
detection of dust around LkCa~14 based on H-band scattered light images,
however, LkCa~14's emission is photospheric out to $\sim$24~{\micron}
\citep{luhman10}, making it suitable for the purposes of this work.

\begin{figure}
\epsscale{1}
\plotone{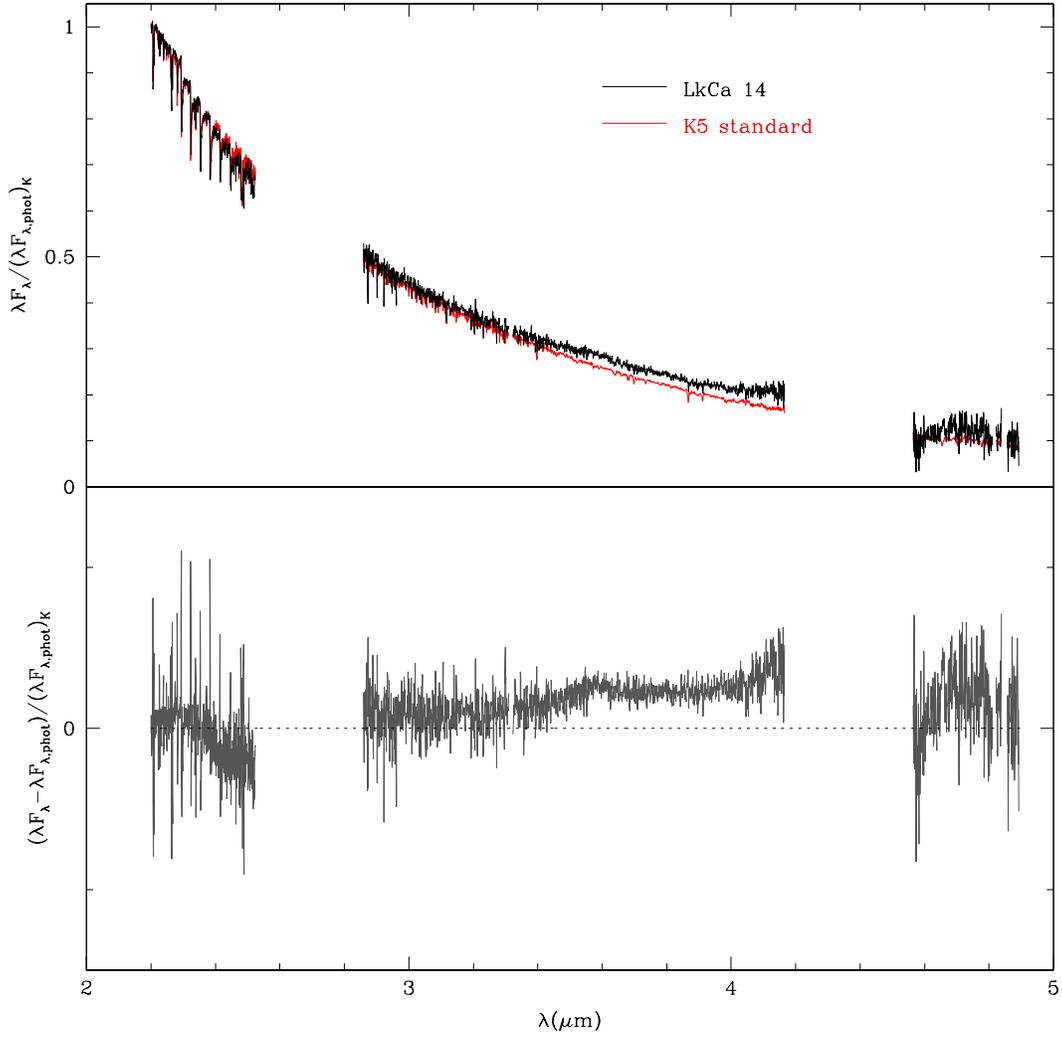}
\caption[Near-infrared emission of the diskless T Tauri star  LkCa~14
relative to a standard star]{Near-infrared emission of the diskless T
Tauri star  LkCa~14 relative to a standard star. The top panel contains
the SpeX spectrum of LkCa~14 (dark line) compared to a K5
standard (light line). LkCa~14 has absorption lines that are not veiled
relative to the standard star and we derive a veiling value (r$_{K}$) of
0 for LkCa~14 between 2.20 -- 2.28~{\micron}. The bottom panel displays
the near-infrared excess emission of LkCa~14. After we subtract the
standard star from the target (bottom), we find that there is no excess emission
above the photosphere of LkCa~14 (gray). The dotted line corresponds to
no excess and the small discrepancy between $\sim$ 3.5--4.2~{\micron}
is likely due to spectral slope uncertainties. This indicates that SpeX
spectra of dwarfs are good templates for the photospheres of young stars.
}
\label{figirtflkca14}
\end{figure}

We measure the amount of veiling in LkCa~14 by comparing its K-band
spectrum to that of a K5 V spectral standard from the IRTF spectral
library (HD36003). We find that LkCa~14 has no veiling
(r$_{K}$=0${\pm}$0.2) from 2.20 --2.28~{\micron}. In order to extract
any excess emission between 2--5~{\micron}, if present, we first
deredden the spectra. The standards from the IRTF library have already
been corrected for reddening. For LkCa~14, we measure a visual
extinction (A$_{V}$) of 0 by matching  V \citep{hog00}, R
\citep{norton07},  and 2MASS photometry to the colors of a K5 stellar
photosphere \citep{kh95} which is scaled to the J-band photometry. We
subtract the standard from the target and find that there is no
significant excess emission at longer wavelengths
(Figure~\ref{figirtflkca14}). There is a small mismatch between the two
spectra from $\sim$3.5--4.2~{\micron} which is most likely due to
uncertainties in the measured spectral slopes of the target.

We conclude that, as expected, LkCa~14 has no veiling and no excess
emission.  This indicates that for the purposes of this work dwarfs in
the IRTF online library are suitable templates.

\subsubsection{LkCa~15} \label{sec:lk}

\citet{espaillat08a} measured a veiling factor at $\sim$2.2 {\micron} of
0.3$\pm$0.2 for LkCa~15.
Using this veiling factor, they extracted the excess emission spectrum
from the SpeX data and fit it with a simple blackbody at a temperature
of 1600~K, corresponding to the behavior expected from the wall of an
inner optically thick disk. This blackbody fit, when taken in
conjunction with the dust clearing inferred from the broad-band SED
\citep{espaillat07b} and millimeter imaging \citep{pietu06}, made LkCa
15 the first confirmed gapped disk around a TTS.

Here we present new SpeX data from $\sim$0.8--5~{\micron} for LkCa~15
and analyze the near-IR excess emission of this object in more detail by
fitting it with the disk wall model of \citet{dalessio05}. We compare
LkCa~15 a K3 V spectral standard  (HD219134) from the IRTF library and
measure an r$_K$ of 0.6$\pm$0.2. While this is in agreement with
\citet{espaillat08a} within the uncertainties, it may also reflect
infrared variability as has been observed in transitional disks
\citep{muzerolle09}. We scaled the spectrum of LkCa~15 relative to the
template according to this veiling factor and dereddened the LkCa~15
spectrum using the \citet{mathis90} dereddening law until it agreed with
the template spectrum at 1~{\micron}.  In this manner, we measure an
extinction of 1.3. Extinction measurements based on matching
V-~, R-~, I-band, and 2MASS photometry to photospheric colors from
\citet{kh95}  derive A$_{V}$=1.7. If we scale
LkCa~15's spectrum using r$_K$=0.4, which is within the veiling
measurement uncertainty, we measure a visual extinction of 1.3 and so
the difference in A$_{V}$ is within the uncertainties of the veiling
determination. Alternatively, this may be a reflection of the fact that
in some cases one can derive higher visual extinctions using color
indices at longer wavelengths than V-R \citep{gullbring98}. 

\begin{figure}
\epsscale{1}
\plotone{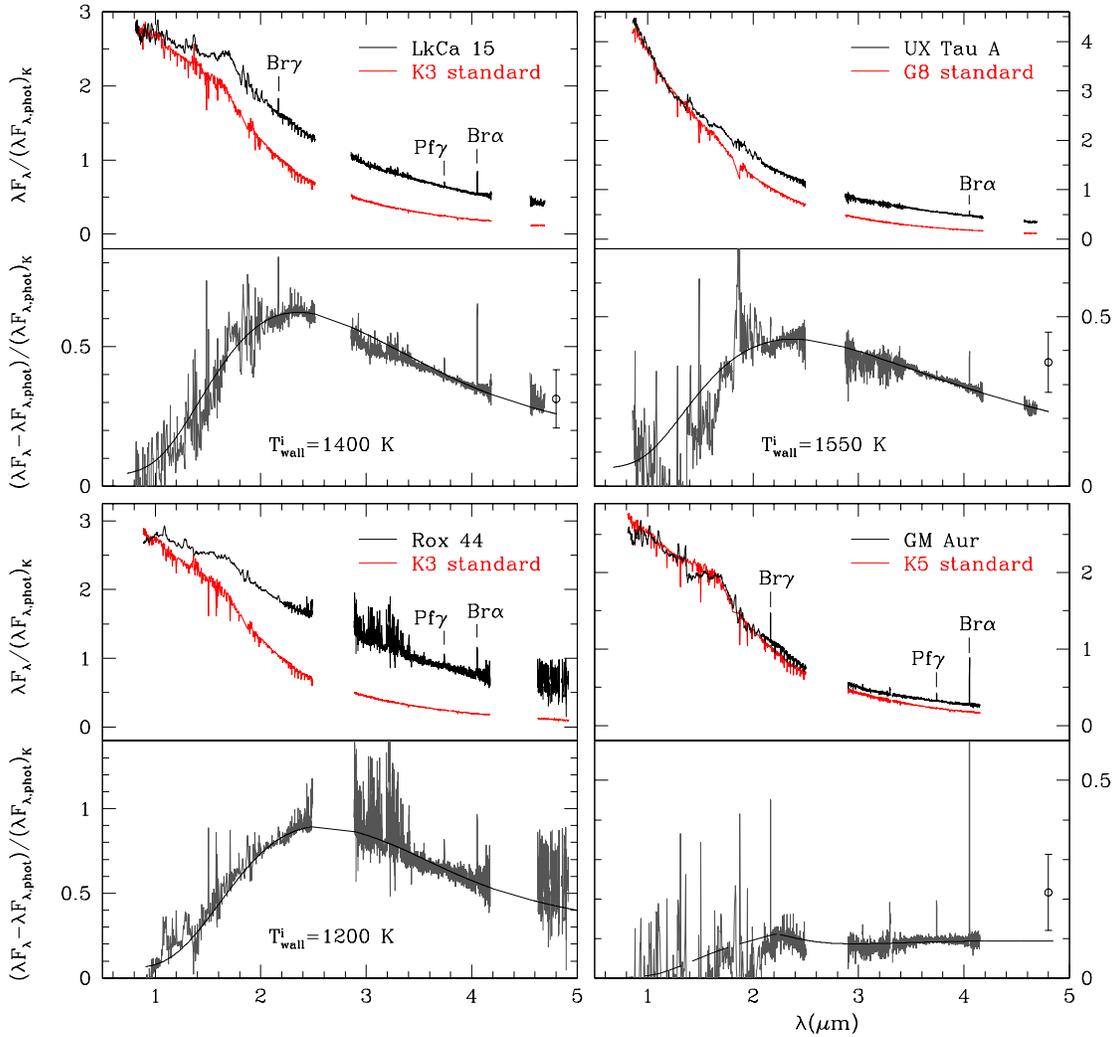}
\caption[SpeX near-infrared emission of the pre-transitional disks of
LkCa~15, UX~Tau~A, and Rox~44 and the transitional disk
of GM~Aur relative to a standard star.]{
SpeX near-infrared emission of the pre-transitional disks of
LkCa~15, UX~Tau~A, and Rox~44 and the transitional disk
of GM~Aur relative to a standard star.  After scaling the targets by $1 + r_{K}$,
we subtracted the standard star spectra (top, light line) from the
veiling-scaled target spectra (top, dark line) to extract the shapes of
the excess emission above the photosphere (bottom, gray).  
We measure an r$_{K}$ of 0.6, 0.4, 0.8, and 0.1 for LkCa~15, UX~Tau~A,
Rox~44, and GM~Aur, respectively.
Additional veiling measurements from \citet{salyk09} at $\sim$4.8~{\micron} (open points) are 
shown.
For the pre-transitional disks, 
we find that the near-infrared excess emission can be
reproduced by the wall of an optically thick inner disk with a temperature
of T$_{wall}^i$ (solid line). 
In the case of GM~Aur,  we can reproduce the near-IR excess emission with an optically
thin dust model (bottom, broken line) composed of sub-micron-sized dust
located within the inner disk hole following \citet{calvet05}.  
We note that GM~Aur was too faint to extract any data at
wavelengths greater than $\sim$4.2~{\micron} and that,
in each of the spectra, 
regions from $\sim$1.2--1.4~{\micron}, $\sim$1.8-1.9~{\micron}, and $\sim$2.8--3.2~{\micron}
are strongly affected by uncertainties in the telluric subtraction.
}
\label{figirtfall}
\end{figure}

After
subtracting the template spectrum from the target spectrum to extract
the excess above the stellar photosphere, we fit the residual excess
continuum with a disk wall model (Figure~\ref{figirtfall}).
For the inner wall model of LkCa~15, we adopt the stellar
properties listed in Table~\ref{tab:prop} which were obtained with
the \citet{siess00} evolutionary tracks.  The mass accretion rate of LkCa~15
 was estimated from its U-band excess following
\citet{gullbring98}.  We adopt a minimum grain size of
0.005~{\micron} as is typically assumed for the interstellar medium
\citep{mathis77} and vary the maximum grain size and the temperature of the optically thin wall
atmosphere (T$_{wall}^i$) to achieve the
best-fit to the SpeX excess emission. We test wall temperatures between
1000--2000~K, corresponding to the range of dust sublimation
temperatures found for a large sample of classical T Tauri stars (CTTS)
and Herbig Ae/Be stars \citep{monnier02}, and maximum grain sizes of
0.25~{\micron}, 1~{\micron}, and 5~{\micron}.  LkCa~15 has significant
10~{\micron} silicate emission, indicative of small grains, and
therefore we do not test larger grain sizes. The best-fit wall to the
excess emission has a temperature of 1400~K, a height of 0.017~AU, and a maximum grain size of
1~{\micron} which, according to Eq.~\ref{rdust}, corresponds to a
wall located $\sim$0.12~AU from the central star
(Figure~\ref{figirtfall}). This detailed model fit is consistent with
the simple 1600~K blackbody fit found in earlier work
\citep{espaillat08a}. The difference in temperature follows from the
additional parameters included in the wall model, namely the effect of
the dust properties and the inclusion of an optically thin wall
atmosphere.  Our detailed fit to the excess continuum is consistent
a veiling measurement at $\sim$4.8~{\micron} by \citet{salyk09} (see
 Figure~\ref{figirtfall}).

\subsubsection{UX~Tau~A} \label{sec:ux}

UX~Tau~A was first modeled as a pre-transitional disk by
\citet{espaillat07b}. To discern the nature of the innermost disk of
this object, here we compare UX Tau~A to a standard star of G8 V
(HD101501) from the IRTF library and find that the absorption lines in
the K-band spectrum of UX~Tau~A are weaker than those in the spectrum of
the standard dwarf between 2.20--2.28~{\micron}.  We measure an r$_K$ of
0.4$\pm$0.2 which is consistent with 2MASS K-band photometry, from which
we infer an r$_K$ of 0.3.

We scale UX~Tau~A's SpeX spectrum relative to the template according to
the derived veiling factor and deredden the target spectrum until it
agrees with the template spectrum at 1~{\micron}.  We measure an
extinction of 2.3 using this method, however, we derive an A$_{V}$ of
1.8 for UX~Tau~A by comparing V-, R-, I-band \citep{kh95} and 2MASS
photometry to a standard G8 photosphere's colors \citep{kh95}. This
difference in A$_{V}$ is within the uncertainties of the veiling
determination.

After extracting the near-infrared excess of UX Tau~A
(Figure~\ref{figirtfall}), we fit it with the wall model described in
{\S}~\ref{sec:mod}. We use the stellar parameters (M$_{*}$, R$_{*}$,
L$_{*}$) listed in Table~\ref{tab:prop} which were derived from the HR
diagram and the \citet{siess00} evolutionary tracks, and we adopt an
inclination for UX~Tau~A of 60$^{\circ}$. As with LkCa~15, we adopt a minimum grain size of
0.005~{\micron} and vary the temperature of the wall between
1000-2000~K.  Since there is no 10~{\micron} silicate feature in
UX~Tau~A, indicating a lack of small grains, we adopt a maximum grain
size of 10~{\micron}.  We cannot discriminate between wall fits using a
maximum grain size of 10~{\microns}, 100~{\microns}, or 1000~{\microns},
however, since the wall's parameters (z$_{wall}^o$, R$_{wall}^o$) do not
vary significantly between the three grain sizes for the purposes of
this work, we adopt the smallest maximum grain size that can reproduce
the SpeX excess spectrum. We find that the best-fit wall has a surface
temperature of $\sim$1550 K, is located 0.15~AU from the star,
and has a height of 0.009~AU
(Figure~\ref{figirtfall}).  Our results are consistent with other
veiling measurements found in the literature.
\citet{edwards06} measure a veiling of 0 at 1~{\micron} and
\citet{salyk09}'s veiling measurement at $\sim$4.8~{\micron}
is consistent with our analysis within the veiling measurement
uncertainties (see Figure~\ref{figirtfall}).

\subsubsection{Rox~44} \label{sec:rox}

Rox~44, located in the Ophiuchus cloud, was identified as a
pre-transitional disk candidate based on its strong 10 {\micron}
silicate emission feature which could not be explained by full disk
models \citep{furlan09,mcclure10}. Rox~44's {\it Spitzer} IRS spectrum
is also strikingly similar in shape to that of LkCa~15. To investigate
the nature of the inner disk of Rox~44, we obtained a near-IR SpeX
spectrum. When compared to a K3 V standard (HD219134) from the IRTF
library,  the absorption lines in the K-band spectrum of Rox~44 are
weaker between 2.20 -- 2.28~{\micron} and we measure an r$_K$ of
0.8$\pm$0.2, consistent with an r$_K$ of 0.7 inferred from 2MASS K-band
photometry.

We scale the spectrum of Rox~44 relative to the template according to
this veiling factor and deredden its spectrum until it agrees with that
of the template spectrum at 1~{\micron} (Figure~\ref{figirtfall}). We
measure a visual extinction of 2.2, in agreement with previous
extinction measurements from \citet{bouvier92}.\footnote{This extinction
differs from the A$_{V}$=3.3 reported by \citet{mcclure10} based on
near-IR colors.  In this work we adopt an A$_{V}$ that agrees with the
optical photometry.  More optical photometry is needed to better determine the extinction
of this object.} The excess emission above the stellar photosphere
can be fit with a wall model (Figure~\ref{figirtfall}) using the stellar
parameters in Table~\ref{tab:prop} and adopting a minimum grain size of
0.005~{\micron}.  We tested maximum grain sizes of 0.25~{\micron},
1~{\micron}, and 5~{\micron} and temperatures between 1000--2000~K. The
best-fit wall to the near-infrared excess continuum of Rox~44 has a height of
0.034~AU, a
maximum grain size of 1 {\micron}, a temperature of 1200~K, and is
located 0.25~AU from the central star (Figure~\ref{figirtfall}).

\subsubsection{GM~Aur} \label{sec:gm}

We now turn to the transitional objects in our sample. The transitional
disk around the single star GM~Aur  \citep[][A. Kraus, private
communication]{pott10, white01} has an inner hole of $\sim$ 20~AU
\citep{hughes09} which contains some sub-micron-sized optically thin
dust \citep{calvet05}. In our veiling analysis, we compare GM~Aur to a
K5V standard (HD36003) and find that GM~Aur's K-band absorption lines
are slightly veiled (r$_K$=0.1${\pm}$0.2) and the object exhibits a
small excess above the stellar photosphere (Figure~\ref{figirtfall}).
This veiling measurement is consistent with a veiling factor of 0.2
inferred from 2MASS K-band photometry.

After scaling the GM~Aur spectrum according to its veiling factor, we
derive an extinction (A$_{V}$) of 1.2 by dereddening its spectrum with
the \citet{mathis90} law until it matches the template spectrum at
1~{\micron}.  However, as in the case of LkCa~15 and UX Tau A, this is
higher than the extinction of 0.8 measured by matching V, R, and I
photometry \citep{kh95}, and 2MASS fluxes to photospheric colors from
\citet{kh95}.

The small excess seen in GM~Aur's SpeX spectrum is in line with the excess
emission above the photosphere seen in its SED from $\sim$
3--8~{\micron} \citep{calvet05}. \citet{calvet05} showed that this
emission comes from optically thin sub-micron-sized dust located within
the inner 5~AU of GM~Aur's 20~AU hole. We can fit the SpeX excess of
GM~Aur with an optically thin dust model (Figure~\ref{figirtfall})
supporting the conclusions of \citet{calvet05}. The optically thin dust
region lies within $\sim$1~AU of the star and is composed of
$\sim$2$\times$10$^{-12}$~{\msun} of dust made up of silicates (47$\%$), organics
(42$\%$), and troilite (11$\%$). The total emission of this optically
thin region is scaled to the vertical optical depth at 10 {\micron}
($\tau_{0}$ $\sim$ 0.021).  We note that since the structure of the dust
within the holes of transitional disks is not known at this time, our
results should be taken as an approximation of how much dust is needed
within the hole to reproduce the SED, not as a detailed fit from which
we can derive the spatial distribution of the optically thin dust near
the star.  High-resolution near-IR interferometry is needed to trace
this component in more detail.

Our results are in agreement with other veiling
measurements at 1~{\micron} \citep{edwards06} and 2.2~{\microns}
\citep{folha99}. 
We note that the
excess emission of GM~Aur derived here does not agree with the J-band veiling measurement of
\citet{folha99}.  However, these authors use a later spectral type standard (K7) in their analysis
which could have resulted in the higher veiling measurement since later type stars
have deeper absorption lines.  Alternatively, this may be a result of variability
as is seen in other transitional disks
\citep{muzerolle09}.

\subsubsection{DM~Tau} \label{sec:dm}

\begin{figure}
\epsscale{1}
\plotone{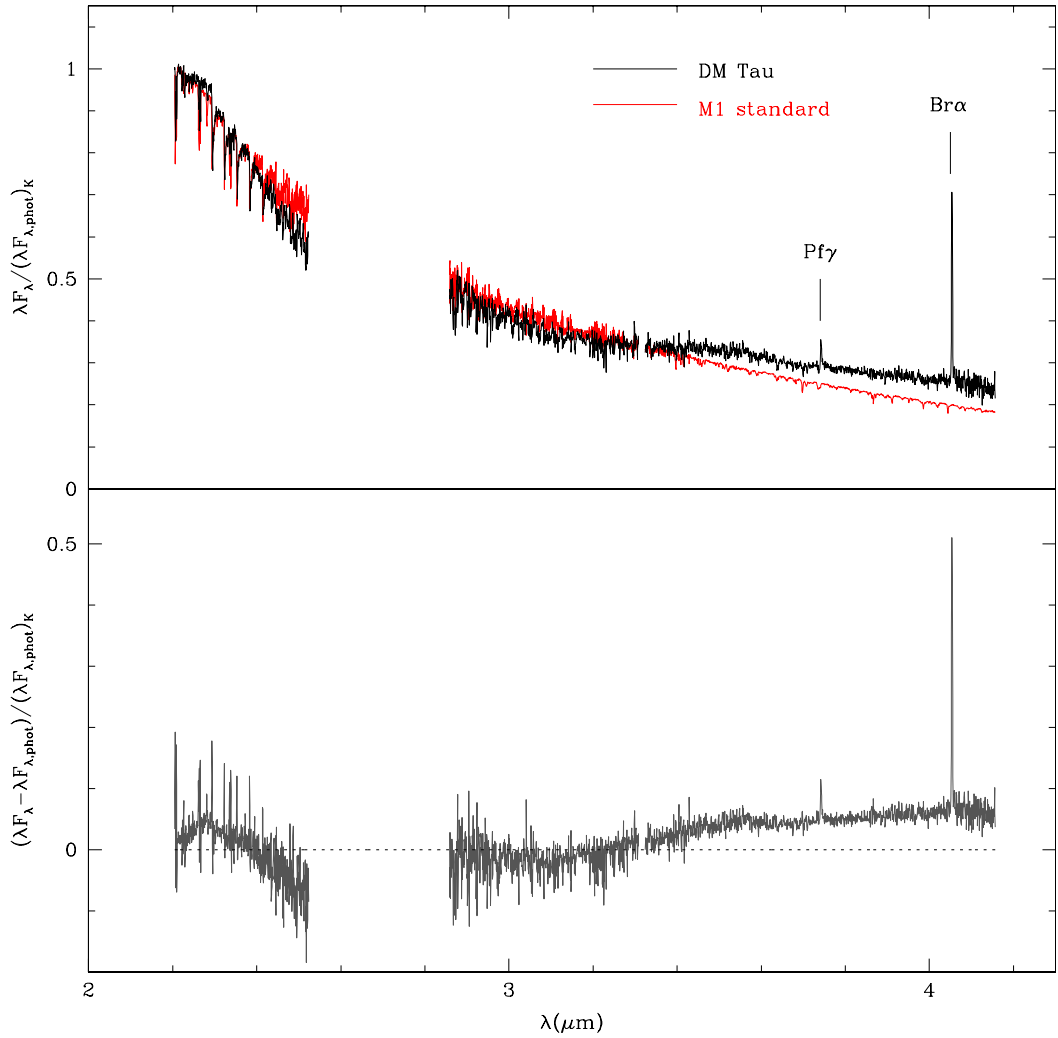}
\caption[Near-infrared emission of the transitional disk of DM~Tau
relative to a standard star]{Near-infrared emission of the transitional
disk of DM~Tau relative to a standard star. In the top panel, we compare
the SpeX spectrum of DM~Tau (solid line) to a M1 dwarf standard
star (broken line).  We derive r$_{K}$=0 for DM~Tau
between 2.20 -- 2.28~{\micron} and find that after subtracting the
standard from the target (bottom panel) there is also no excess at
longer wavelengths. The dotted line in the bottom panel corresponds to
zero excess. The small discrepancy between $\sim$ 2.4--2.5~{\micron} is
likely due to spectral slope uncertainties. 
Note that DM~Tau was too faint to extract any data at
wavelengths greater than $\sim$ 4.2~{\micron}.
}
\label{figirtfdmtau}
\end{figure}

DM~Tau is a single star \citep[][A. Kraus, private
communication]{pott10,white01} surrounded by a transitional disk with a
$\sim$3~AU hole that is empty of small dust grains \citep{calvet05}. The
absorption lines of DM~Tau are not veiled relative to a M1 V standard
(HD42581; Figure~\ref{figirtfdmtau}) and we measure an r$_{K}$ of
0${\pm}$0.2 between 2.20 -- 2.28~{\micron} in the SpeX spectrum.

We derive an extinction of (A$_{V}$) of 0.7 for DM~Tau by fitting V-,
R-, I-band photometry  \citep{kh95} and 2MASS colors to an M1 stellar
photosphere from \citet{kh95}.  The spectrum is dereddened with this
extinction and the \citet{mathis90} dereddening law.  We note that we
did not obtain a low-resolution SpeX spectrum at shorter wavelengths for
this target.  Subtraction of the standard from the target reveals that
DM~Tau exhibits no significant near-IR excess emission at 2--5~{\micron} 
(Figure~\ref{figirtfdmtau}), in agreement with its lack of
emission above the photosphere in the K-band (see SED in Calvet et
al.~2005 or Furlan et al.~2006) as well as a veiling measurement
at $\sim$4.8~{\micron} that is consistent with no excess emission at this wavelength
\citep{salyk09}. 
As in the case of LkCa~14, DM~Tau shows
a small mismatch when compared to the standard, which may be an
indication of spectral slope uncertainties. The lack of veiling and
near-IR excess emission in DM~Tau is consistent with the fact that this
object has no excess emission above the photosphere from the K-band up
to $\sim$ 8~{\micron} \citep{calvet05}, indicating that DM~Tau has relatively no
small dust within its hole.

\subsection{Broad-Band SED Modeling of Pre-Transitional Disks} \label{sec:models}

Here we model the broad-band SEDs of the pre-transitional disks of
LkCa~15, UX~Tau~A, and Rox~44 with the best-fit inner wall model
from {\S}~\ref{sec:nearir},
including the shadowing of this inner wall on the outer disk.
LkCa~15
and UX~Tau~A have been modeled previously \citep{espaillat07b},
but here we incorporate their newly derived stellar properties (\S~\ref{sec:spt}),
and inner walls ({\S}~\ref{sec:lk} and {\S}~\ref{sec:ux}), and a finite star.

\subsubsection{LkCa~15}
 
\begin{figure}
\epsscale{1}
\plotone{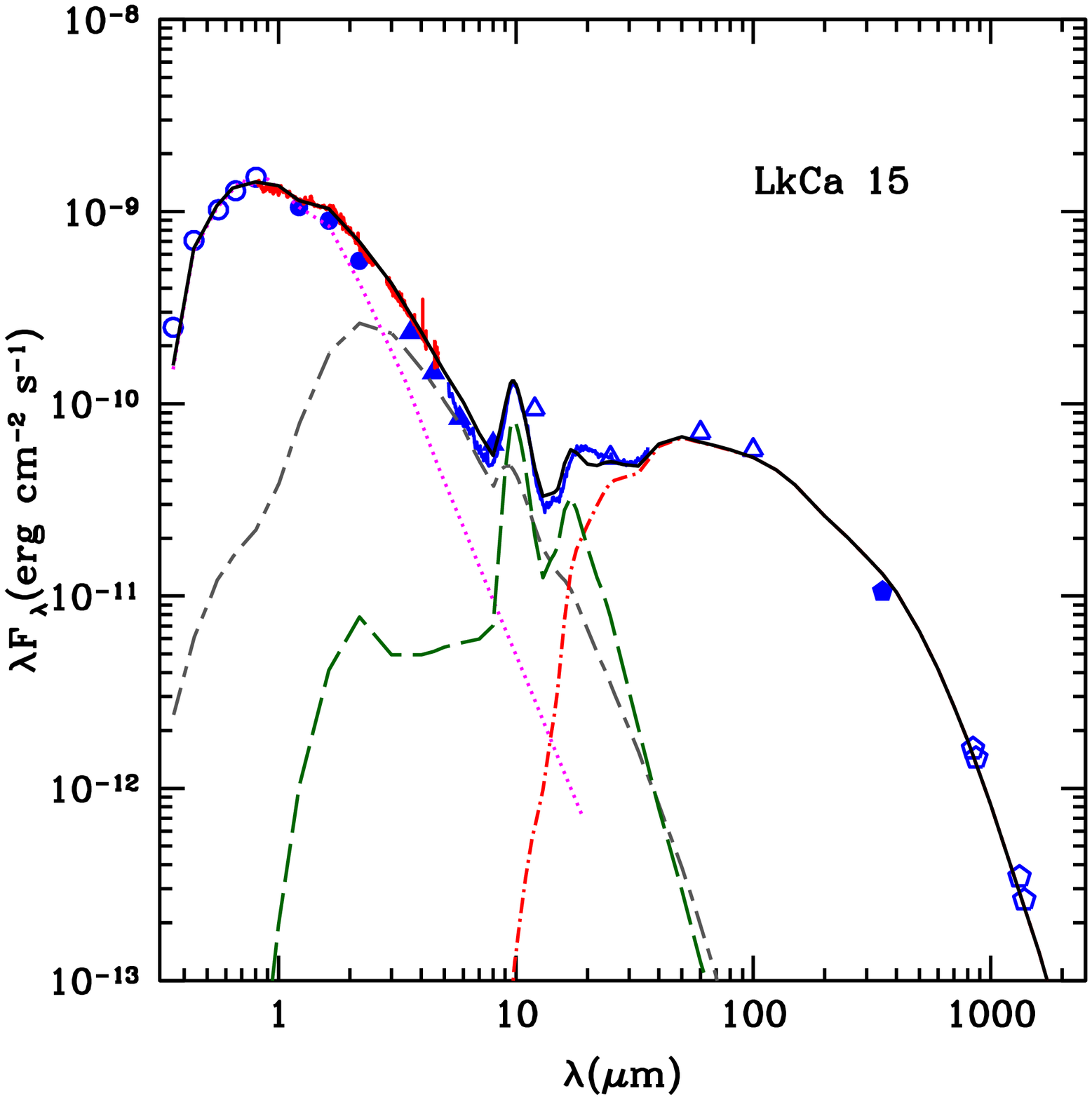}
\caption[SED and pre-transitional disk model of
LkCa~15]{Pre-transitional disk model of LkCa~15. The best-fit model to
LkCa~15 (solid black line) consists of an inner optically thick disk
which extends from the dust destruction radius to $<$0.26~AU and an
outer disk from 58 -- 300~AU. Within the inner 4~AU of the gap between
the inner and outer disk, there is $\sim$10$^{-11}$~{\msun} of ISM-sized
optically thin dust. Separate model components are the stellar
photosphere \citep[magenta dotted line;][]{kh95}, the inner disk (gray
short-long-dash), the outer disk (red dot-short-dash), and the optically
thin small dust located within the disk gap (green long-dash).
[See the electronic edition of the Journal for a color version of this
figure.]
}
\label{figmodellkca15}
\end{figure}

For LkCa~15, we measure a 58~AU
gap that contains some small optically thin dust.  This is compatible with
millimeter interferometric imaging by \citet{pietu06} which finds a
cavity of 46~AU in this disk, especially when one takes into account
the uncertainties introduced by dust opacity one assumes (see ${\S}$~\ref{sec:cav}).  
Figure~\ref{figmodellkca15} contains the broad-band SED of LkCa~15.
Observations include ground-based optical photometry \citep[open
circles;][]{kh95}, and J, H, K (filled circles; 2MASS), SpeX (red solid
line; this work), $Spitzer$ IRAC \citep[blue solid
triangles;][]{luhman10}, $Spitzer$ IRS \citep[blue solid
line;][]{furlan06}, IRAS \citep[open triangles;][]{weaver92}, and
millimeter (filled pentagon from Andrews \& Williams 2005; open
pentagons from Espaillat et al. 2007) data.  The SpeX spectrum used in
this paper is not flux calibrated; here we scale it to the adopted
stellar photosphere \citep{kh95} using the veiling derived in
{\S}~\ref{sec:lk}

In Figure~\ref{figmodellkca15}, we use the best-fit wall model found in
{\S}~\ref{sec:lk}.  We note that while this wall is a good fit to the SpeX spectrum,
it over-predicts the {\it Spitzer} IRAC and IRS data.  This can be explained
by the uncertainties inherent in our veiling measurement or the fact that infrared variability in these
types of disks is common \citep[Espaillat et al., in preparation;][]{muzerolle09}.
The 10 {\micron} silicate emission can be
fit by an optically thin dust model which contains
$\sim$2$\times$10$^{-11}$~${\msun}$ of optically thin dust within the
inner 4~AU of the gap.  This region is composed of 79$\%$ silicates,
12$\%$ organics, and 9$\%$ troilite and the total emission is scaled to
the vertical optical depth at 10 {\micron} ($\tau_{0}$ $\sim$ 0.009).
The outer disk has ${\alpha}$=0.0005 and ${\epsilon}$=0.001 and is
inwardly truncated at about 58~AU. The inner wall dominates the near-IR
emission and we constrain the inner disk to $<$.19~AU in radius,
assuming that the inner disk has the same $\alpha$ and $\epsilon$ as the
outer disk.  Model parameters are listed in Table~\ref{tab:prop}.  

Our results for LkCa~15 differ from \citet{espaillat07b}. In first
modeling the outer disk of LkCa~15, \citet{espaillat07b} adopted a
spectral type of K5 (T$_{*}$=4350~K) for this object following
\citet{herbig86}. However, the spectral type we
derived in {\S}~\ref{sec:spt} for LkCa~15 (K3) corresponds to a
higher temperature (T$_{*}$=4730~K) than previously
adopted. The largest difference between the disk model presented here
and that in \citet{espaillat07b} is the location of the outer disk wall.
Assuming a spectral type of K5 the disk wall was located at 46~AU while
for a K3 star the wall is located at 58~AU.  This difference is expected
since the location of the wall, given by Eq.~\ref{walleqn}, depends
on $L_* $ and $L_{acc}$ and these parameters are now higher with respect
to the previous model.

\begin{figure}
\epsscale{1}
\plotone{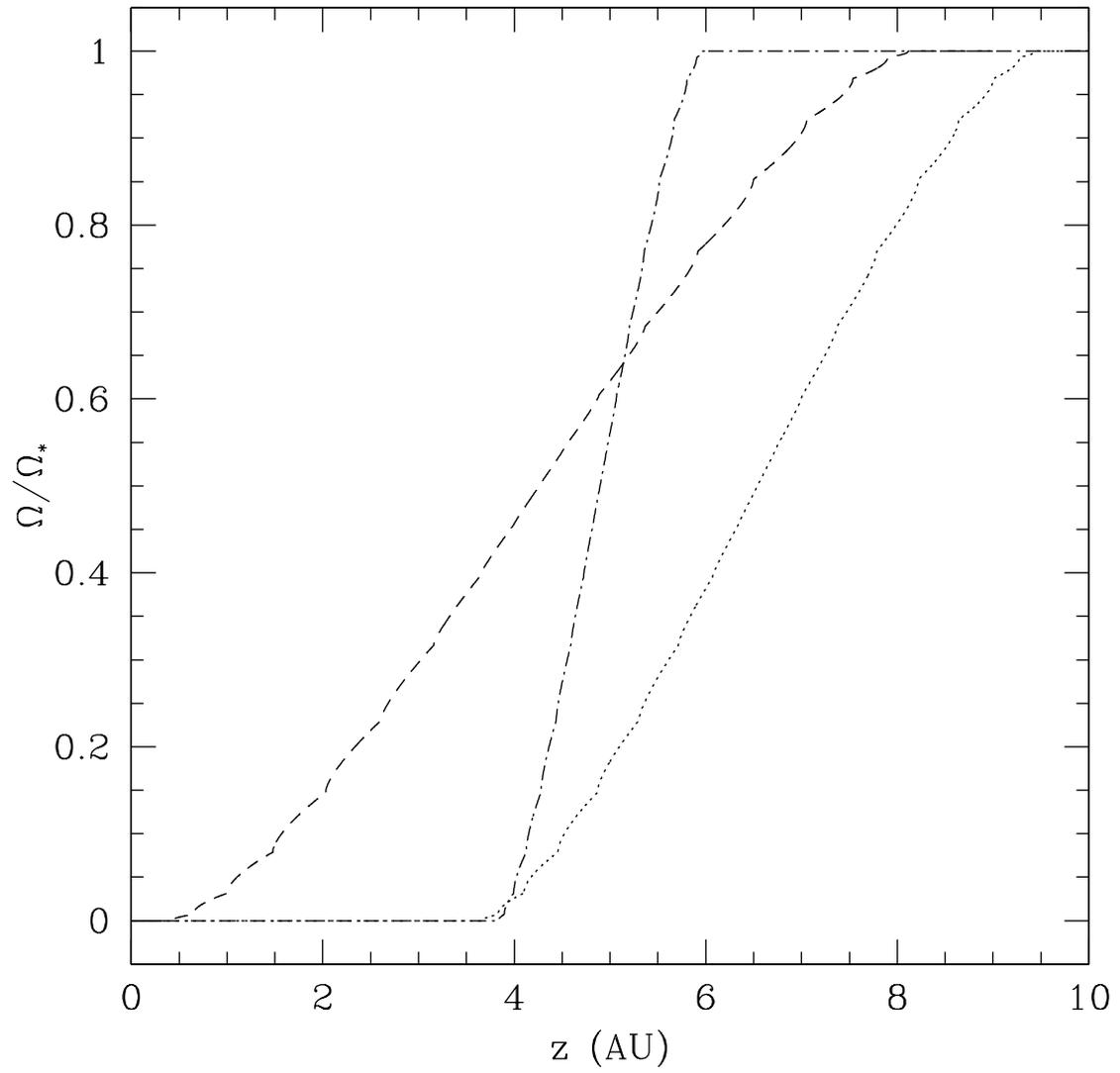}
\caption[Solid angle of LkCa~15, UX~Tau~A, and Rox~44.] {Solid angle of
LkCa~15 (dotted line), UX~Tau~A (dashed line), and Rox~44 (dot- dashed
line) as seen from different heights ($z$) at the radius of the outer wall.
$\Omega$ is given in units of $\Omega_*$, the solid angle of the entire star.}
\label{shadow3}
\end{figure}

Based on the equations presented in the Appendix, the inner wall of
LkCa~15 would cast an umbra of about 7~AU on the outer wall (measured from the midplane) 
in the case
that the star is taken to be a point source. If the star is taken to be
a finite source, LkCa~15's outer wall has an umbra of $\sim$4~AU
and the upper edge of the penumbra is located at $\sim$9~AU
(Eq.~\ref{eqn:zp}~\&~\ref{eqn:zn}). The result for the solid angle
(Eq.~\ref{eqn:omega}) of the star seen by an observer at different
heights of the outer wall of LkCa~15 is shown in Figure~\ref{shadow3}.
Taking into account the shadowing of a finite star, the outer wall of
LkCa~15 is 12.9~AU high.  This is 2.6~H, where $H$ is the gas scale
height, and is consistent within 15$\%$ of the independently calculated
height of the outer disk (11.7~AU) at $\sim$58~AU.  Therefore, while the
outer wall is partially shadowed by the inner wall, we conclude that the
outer disk behind the wall is not completely
shadowed by the inner disk. 
The height
of outer wall in this work is higher than previously reported by
\citet{espaillat07b}.  The z$_{wall}$ in that paper represents the part
of the wall that is illuminated when the star is taken to be a point
source, but was mistakenly defined as the height of the wall above the
midplane.

\subsubsection{UX~Tau~A}

\begin{figure}
\epsscale{1}
\plotone{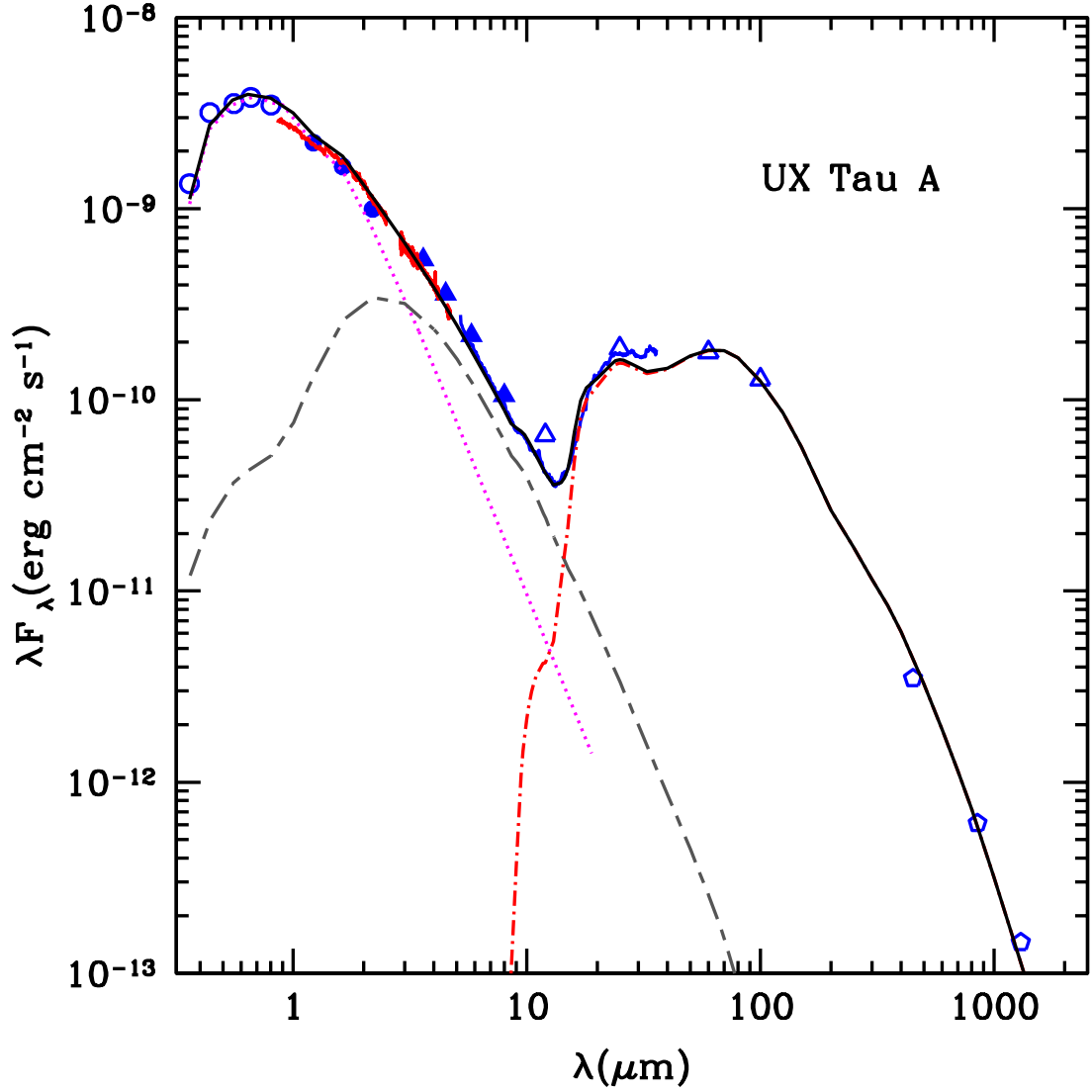}
\caption[SED and pre-transitional disk model of UX Tau
A]{Pre-transitional disk model of UX Tau~A. UX Tau~A's best-fitting model
 (solid black line) has an inner optically thick disk which
extends from the dust destruction radius to $<$0.21~AU and an outer disk
from 71 -- 300~AU. The gap between the inner and outer disk is
relatively empty of small dust grains. Separate model components are the
same as used in Figure~\ref{figmodellkca15}. [See the electronic edition
of the Journal for a color version of this figure.]
}
\label{figuxtauamodel}
\end{figure}

For UX~Tau~A, we find a ~71~AU gap that is relatively devoid of small
dust grains, as indicated by its lack of a 10 {\micron} silicate feature
(Figure~\ref{figuxtauamodel}). Observations shown in
Figure~\ref{figuxtauamodel}  include ground-based optical photometry
\citep[open circles;][]{kh95}, J, H, K photometry (filled circles;
2MASS), a SpeX spectrum (solid red line; this work), $Spitzer$ IRAC
photometry  \citep[blue solid triangles;][]{luhman10}, a $Spitzer$ IRS
spectrum \citep[blue solid line;][]{furlan06}, IRAS photometry
\citep[open triangles;][]{weaver92}, and millimeter data
\citep[pentagons;][]{andrews05}.  Our SpeX spectrum is not
flux-calibrated and we scale it to the adopted stellar photosphere using
the veiling derived in {\S}~\ref{sec:ux}.

We use the best-fit wall model found
in {\S}~\ref{sec:ux} in fitting the SED emission presented in Figure~\ref{figuxtauamodel}. 
We find that the outer disk has
${\alpha}$=0.004 and ${\epsilon}$=0.001 and is inwardly truncated at
about 71~AU. The inner wall dominates the near-IR emission and we
constrain the inner disk, assuming it has the same settling and
viscosity as the outer disk, to $<$.21~AU in radius.  Model parameters
are listed in Table~\ref{tab:prop}.

The outer wall of UX~Tau~A would have an umbra of about 4~AU in the case
that the star is taken to be a point source. If the star is taken to be
a finite source, the size of the umbra is $\sim$0.4~AU and the upper limit
of the penumbra is $\sim$8~AU. The result for the solid angle seen by an
observer at different heights of the outer wall of UX~Tau~A is shown in
Figure~\ref{shadow3}. The outer wall of UX~Tau~A is 13.8~AU (2.0~H) high
and is consistent within 15$\%$ of the height of the outer disk
(12.5~AU) at $\sim$71~AU, which is calculated independently.  The fully
illuminated part of the outer wall is at a temperature where ice has
sublimated (110~K), however, in the 
partially
shadowed portions of the wall the
temperature is lower. In order to make the composition of the wall
consistent,  we remove ice from the 
partially
shadowed portions of the wall. We
note that UX~Tau~A has crystalline silicates \citep{espaillat07b}, which
we do not include here and would contribute to the IRS spectrum between
$\sim$20 -- 35 {\microns}.

As in the case of LkCa~15, our results for UX~Tau~A differ from \citet{espaillat07b}
because we are adopting a different spectral type in this paper.  \citet{espaillat07b}'s model
was based on a
spectral type of K2 (T$_{*}$=4900~K) following
\citet{herbig77} and \citet{cohen79}.  The spectral type we
derived in {\S}~\ref{sec:spt} for UX~Tau~A (G8) corresponds to a
significantly higher temperature (T$_{*}$=5520~K) than previously
adopted.  The difference in R$_{wall}^o$ is expected following Eq.~\ref{walleqn}.

\subsubsection{Rox~44}

\begin{figure}
\epsscale{1}
\plotone{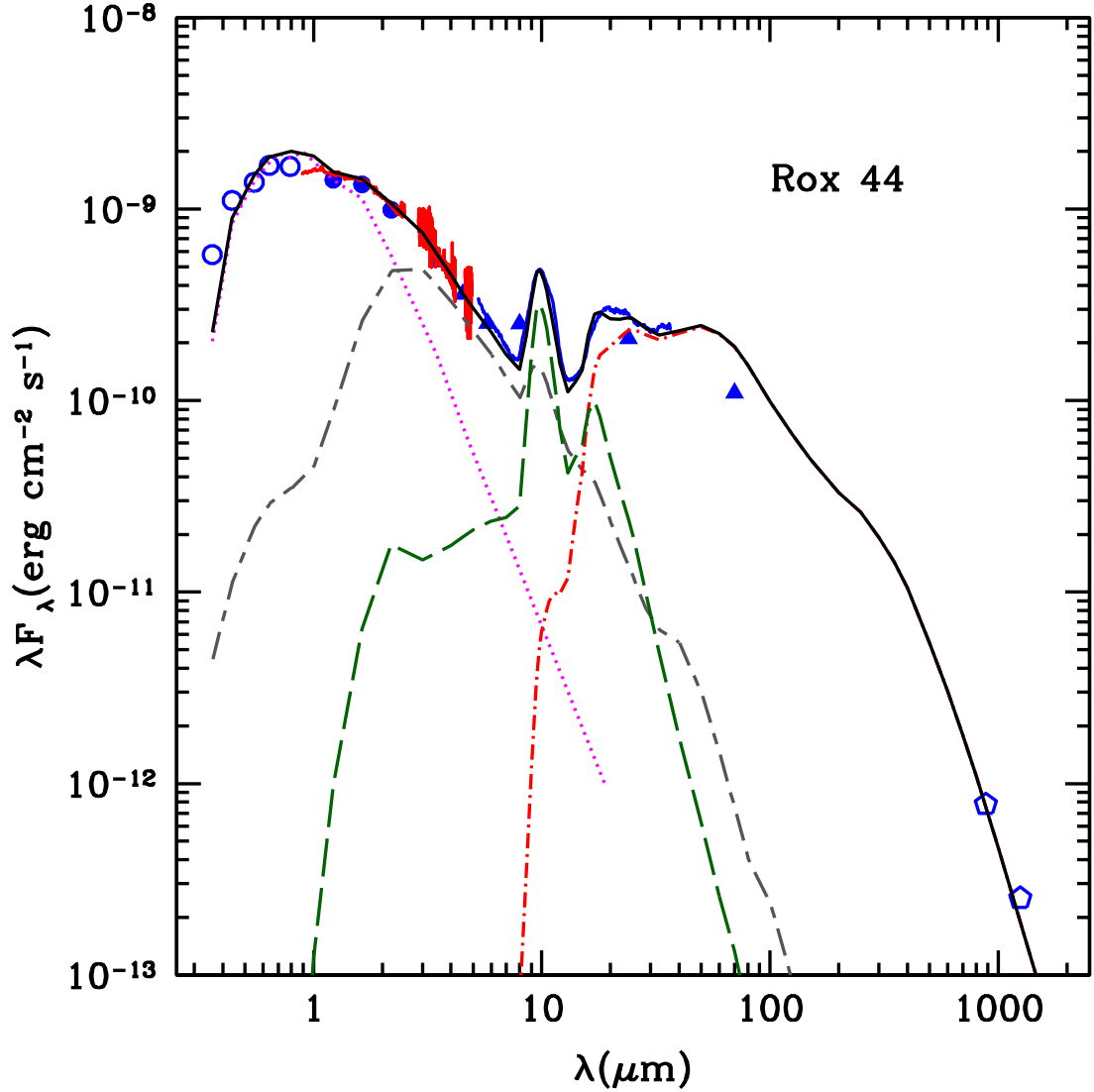}
\caption[SED and pre-transitional disk model of Rox~44]{Pre-transitional
disk model of Rox~44. The best-fit model to Rox~44 (solid black line)
consists of an inner optically thick disk which extends from the dust
destruction radius to $<$0.4~AU and an outer disk from 36 -- 300~AU.
Within the inner 2~AU of the gap between the inner and outer disk, there
is $\sim$10$^{-11}$~{\msun} of ISM-sized optically thin dust. Separate
model components are the same as used in Figure~\ref{figmodellkca15}.
[See the electronic edition of the Journal for a color version of this
figure.]
}
\label{figroxmodel}
\end{figure}

Rox~44 has a 36~AU gap that contains some ISM-sized optically thin dust.
The size of this gap is consistent with the 33~AU cavity observed
with the {\it SMA} \citep{andrews09}.
Observations shown in Figure~\ref{figroxmodel} include ground-based
optical \citep[open circles;][]{herbst94}, J, H, K (filled circles;
2MASS), SpeX (solid red line; this work), $Spitzer$ IRAC and MIPS
\citep[blue solid triangles]{evans09,padgett08}, $Spitzer$ IRS
\citep[blue solid line;][]{furlan06}, and millimeter
\citep[pentagons;][]{andrews09,nuernberger98} data. The SpeX spectrum is
scaled to the adopted stellar photosphere using the veiling
derived in {\S}~\ref{sec:rox}.

We use the best-fit wall model found in {\S}~\ref{sec:rox} in the SED model
fit presented in Figure~\ref{figroxmodel}. 
Rox~44 has $\sim$2$\times$10$^{-11}$~${\msun}$ of
optically thin dust within 2~AU that contributes to the 10 {\micron}
silicate feature and that is composed of 79$\%$ silicates, 12$\%$ organics,
and 9$\%$ troilite.  The total emission of this region is scaled to the
vertical optical depth at 10 {\micron}, $\tau_{0}$ $\sim$ 0.036. The
outer disk has ${\alpha}$=0.006 and ${\epsilon}$=0.01 and is inwardly
truncated at about 36~AU. The inner wall dominates the near-IR emission
and we constrain the inner disk to $<$.4~AU in radius.  Model parameters
are listed in Table~\ref{tab:prop}.

In the case that the star is taken to be a point source, the outer wall
of Rox~44 would have an umbra of $\sim$5~AU. For a finite-source star,
the size of the umbra is $\sim$4~AU and the upper edge of the penumbra
is located at $\sim$6~AU (see Figure~\ref{shadow3}). Taking into
consideration this shadowing of the outer disk, the outer wall of Rox~44
has z$_{wall}^o$=9.9~AU (3.5~H) and this height is consistent within
$\sim$35$\%$ with the height of the outer disk at $\sim$36~AU (7.3~AU).
We note that far-infrared variability is likely in this object given that the {\it Spitzer}
IRS and MIPS data were taken at different epochs and are not in agreement, even
when taking the uncertainties of the observations into consideration.  If the far-infrared flux
of Rox~44 was less than is seen in Figure~\ref{figroxmodel}
 at the time when the SpeX observations were taken, then we would be overestimating the
 height of the outer wall.  Simultaneous observations of this object in the near- and far-infrared
 would be useful in testing this.  Additional far-infrared and sub-millimeter data
 would also help further constrain the properties of the outer disk.

\section{Discussion} \label{sec:discuss}

\subsection{Pre-Transitional Disk Structure} \label{sec:ptd}

It is now accepted that full disks have a sharp transition at the dust
destruction radius, inside of which the temperature is too high for dust to exist.
This transition appears as a ``wall'' that is  frontally illuminated by the star,
since for typical mass accretion rates the inner disk gas is 
optically thin \citep{muzerolle04}.
\citet{muzerolle03} found that nine T Tauri stars surrounded by full
disks in Taurus had SpeX near-IR excess emission which could be fit by
single temperature blackbodies with temperatures that fell within the
range of dust sublimation temperatures (1000--2000 K) found for a larger
sample by \citet{monnier02}. \citet{muzerolle03}'s fits are evidence
that there is optically thick material located at the dust destruction
radius in these full disks, most likely from the inner wall of the disk
which emits primarily in the near-IR.

Pre-transitional disks exhibit the same behavior as full disks in the
near-IR.  In this work, we can fit the continua between 1--5~{\micron}
of the pre-transitional disks in our sample with an inner disk wall
which is located at the dust destruction temperature. This conclusion is
consistent with independent veiling measurements found in the literature
(see {\S}~\ref{sec:nearir}) as well as near-IR interferometric
measurements.  Using the Keck interferometer, \citet{pott10} find that
the K-band emission of LkCa~15 and UX~Tau~A originates from very small
radii.  This is in good agreement with what we report in this work
(Table~\ref{tab:prop}) if one takes into account that \citet{pott10}
used simple models assuming a face-on disk and the derived radii depend
on the dust opacities that are adopted.  Independent SED modeling also
supports that the inner disk of LkCa~15 is optically thick
\citep{mulders10}. We note that \citet{mulders10} claim that the
optically thick inner disk of LkCa~15 extends out to 1~AU based on
fitting broad-band photometry and assuming that the shadowing in the
disk is due to a point source.  This radius is inconsistent with the
results of \citet{pott10} and we find that an inner disk of this size
does not fit the {\it Spitzer} IRS spectrum.

While pre-transitional disks have the same behavior as full disks in the
near-IR,  at longer wavelengths in their SEDs and in millimeter images,
they display evidence for large dust clearings within the disk.  In the
{\it Spitzer} IRS spectra of LkCa~15, UX~Tau~A, and Rox~44, we see a
pronounced decrease in the emission at $\sim$20{\micron} which indicates
a drop in the opacity of the disk, most likely due to the removal of
dust.  This is supported by millimeter interferometric observations of
LkCa~15 \citep{pietu06}, UX~Tau~A (S. Andrews, private communication),
and Rox~44 \citep{andrews09} which see cavities in the images and sharp
drops in the visibilities, evidence that dust in the disk  has been
removed.  In conjunction with the evidence for an optically thick inner
disk, this leads one to the conclusion that pre-transitional disks have
gaps in their disks.

\begin{figure}
\epsscale{1.2}
\plotone{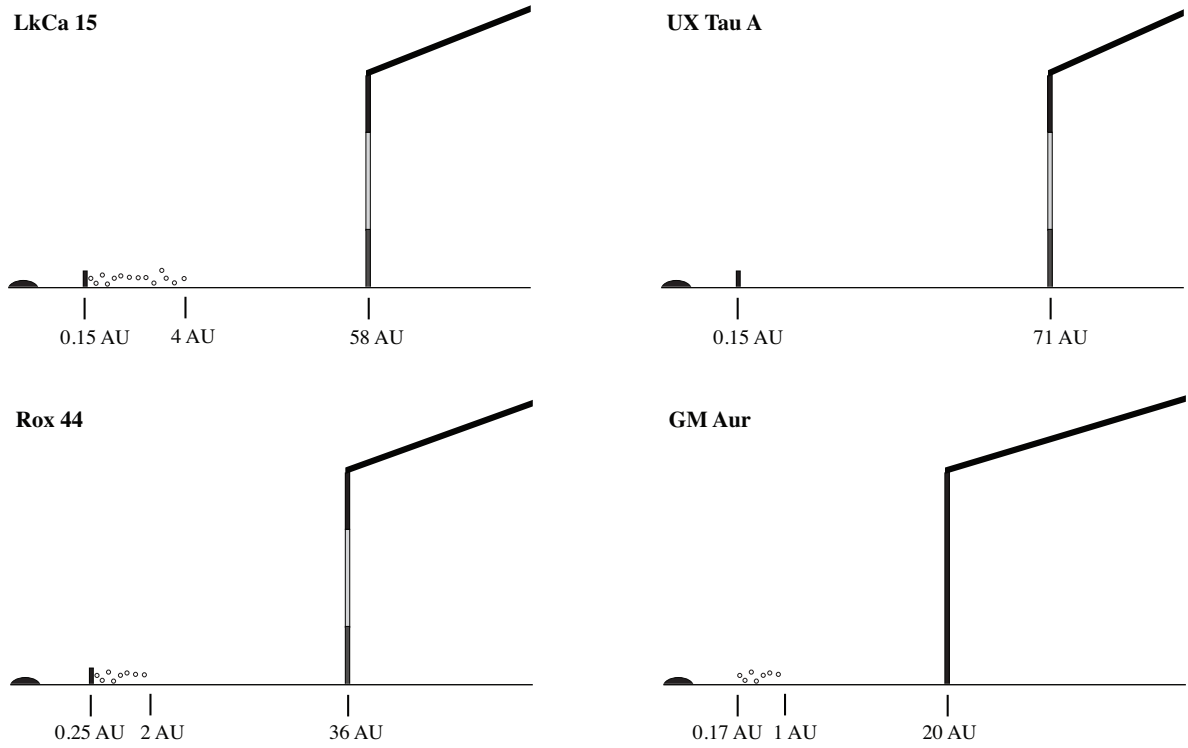}
\caption[Schematic illustration of the
pre-transitional disks of LkCa~15, UX~Tau~A, and Rox~44 and the transitional disk
of GM~Aur] {Schematic illustration of the
pre-transitional disks of LkCa~15, UX~Tau~A, and Rox~44 and the transitional disk
of GM~Aur.  The black semicircles represent the central stars and the small, empty circles
are ISM-sized optically thin dust.  The vertical lines are the disk walls.  For the wall of the
outer disk, black represents the portion of the wall that is fully illuminated by the star,
light gray corresponds to the part of the wall that is in the penumbra of the inner disk, and dark gray is 
the portion of the wall that is
in the umbra.  In each case, the surface of the outer disk is fully illuminated.  The
transitional disk of DM Tau,
which is not shown here, has an inner disk hole similar to what is seen in GM Aur, but it is devoid of optically thin small dust.
Structures here are not drawn to scale.}
\label{schematic}
\end{figure}

These gaps in pre-transitional disks are clearly different from the
inner holes detected in transitional disks (see Figure~\ref{schematic}
for a schematic). DM~Tau has an inner hole of 3~AU which is relatively
clear of small dust as indicated by the lack of near-infrared emission
in its {\it Spitzer} IRS spectrum \citep{calvet05}. This is supported by
the lack of veiling in the K-band as well as photospheric near-infrared
emission, resembling what is seen in the diskless LkCa~14, indicating
that the inner disk of DM Tau is devoid of small dust. GM~Aur is another
transitional disk studied in this paper and its SED indicates that it
has a $\sim$20~AU hole \citep{calvet05} which has been confirmed in the
millimeter \citep{hughes09}. GM~Aur has a small excess above the
photosphere in the near-infrared, as measured by its {\it Spitzer} IRS
spectrum, which has been interpreted as originating from optically thin
dust located within the disk hole \citep{calvet05}.  In this work we
demonstrate that the small, flat near-IR excess continuum extracted from
the SpeX spectrum can be fit by a model with $\sim$10$^{-12}$~{\msun} of
optically thin dust within 1~AU of the star. About 85$\%$ of the
emission in the K-band originates in the inner 0.5~AU of the disk,
consistent with near-interferometric measurements \citep{pott10}.

An alternative model for the inner disk of pre-transitional disks
proposed by \citet{espaillat07b} and further investigated by
\citet{mulders10} is that the inner disks of these objects are optically
thin. However, GM~Aur's near-IR emission is strikingly different from
that seen in our pre-transitional disk sample.   GM~Aur's near-IR excess
is relatively flat, reflecting a range of different temperatures. On the
other hand, the near-IR excess emission of pre-transitional disks
resembles a single temperature blackbody. Optically thin dust spread out
over several radii will not appear as a single temperature blackbody and
if one confines this optically thin dust to very small radii, one would
need a shell of dust engulfing the star \citep{mulders10} for which
there is yet no supporting evidence.

\begin{figure}
\epsscale{1}
\plotone{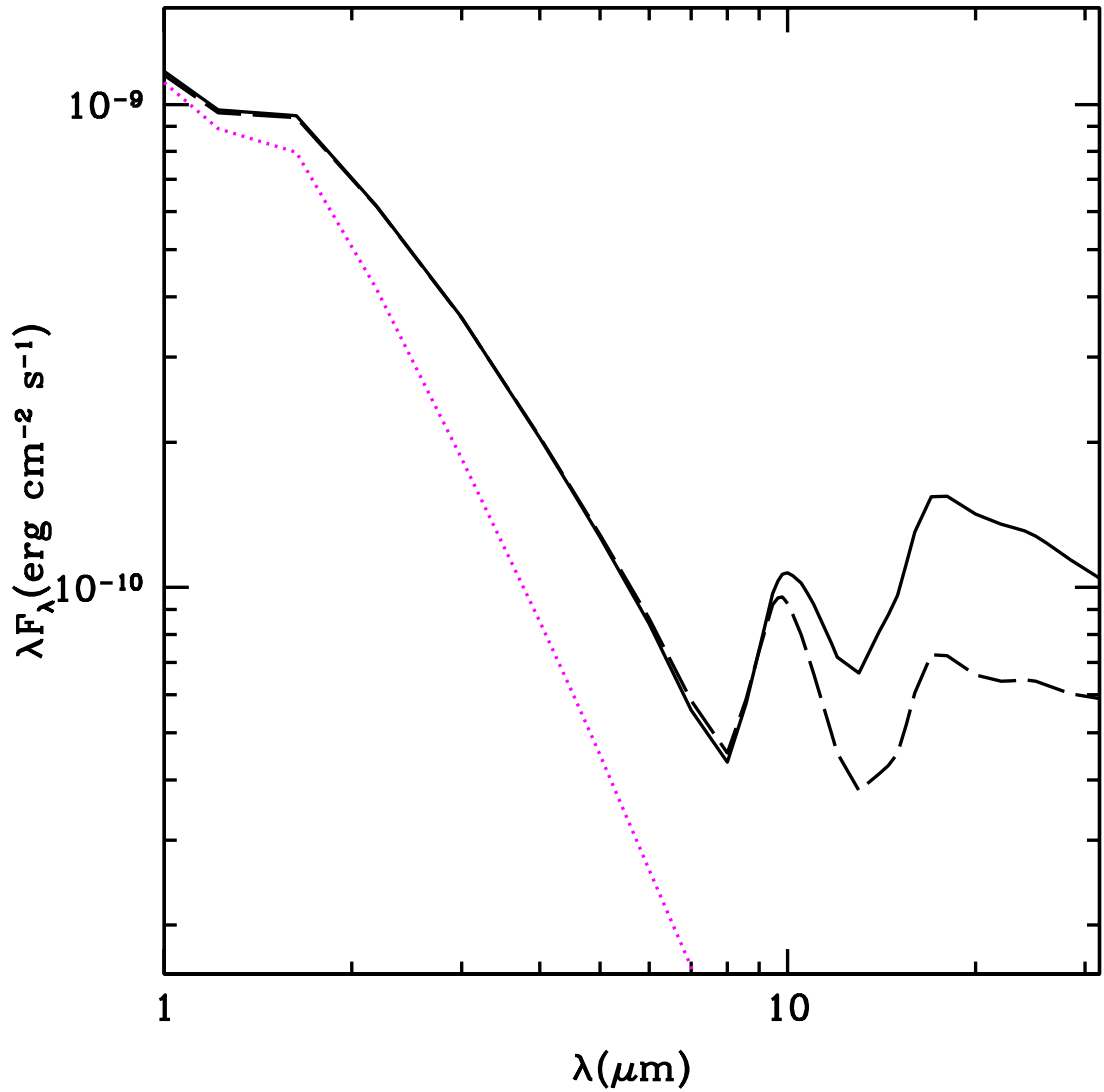}
\caption[Simulation of a gapped disk compared to a typical full disk model]{
Simulation of a gapped disk compared to a typical full disk model.  A pre-transitional disk with a gap extending from $\sim$0.3~AU to 4~AU (solid line) has a SED that diverges from a full disk SED (dashed line) beyond 10~{\micron}.  This
is due to the contribution of the illuminated portion of the outer wall in the gapped disk model.
The dotted line corresponds to the stellar photosphere.  }
\label{gaptest}
\end{figure}

Since the inner disk in pre-transitional disks is optically thick, it
will cast a shadow on the outer disk. The shadow cast by the inner wall
on the outer wall has significant consequences on the sizes of gaps that
we can detect with {\it Spitzer}.  The majority of the emission traced
by IRS comes from within the inner 10~AU of the disk and over 80$\%$ of
the emission at 10~{\micron} comes from within 1~AU \citep{dalessio06,
espaillat09}. Therefore, the {\it Spitzer} IRS instrument will be most
sensitive to clearings in which some of the dust located at radii
$<$1~AU has been removed. To test the limits of the smallest gap IRS
could detect  around a K-type star, we simulated gaps in a disk around a
0.5~$\msun$ star with a mass accretion rate of
10$^{-8}$~$\msun$~yr$^{-1}$, an $\epsilon$ of 0.01, and an inclination
of 60$^{\circ}$. In this disk, we find that the smallest gap that will
cause a noticeable ``dip'' in the {\it Spitzer} spectrum will be
$\sim$4~AU. As is seen in Figure~\ref{gaptest}, the SED of such a gapped
disk begins to differ from that of a full disk beyond 10~{\micron}. This
is because, relative to a full disk, there is extra emission from the
wall of the outer disk. Since the height of the disk increases with
radius, the inner disk can only  extend from the dust destruction radius
(0.12~AU) out to about 0.3~AU before it significantly shadows the outer wall,
thereby dimming the emission of the outer wall and making it
indistinguishable from a full disk. Furthermore, IRS cannot easily
detect gaps whose inner boundary is outside of 1~AU \citep[e.g.~a gap
spanning  5--10~AU in the disk;][]{espaillat09} and it will be up to
{\it ALMA} to detect such gaps.

There are some caveats to keep in mind regarding the above estimate of
the IRS gap detection limit. 
First, we are assuming that the inner disk has the same amount of dust
settling as the outer disk.  However, if the inner disk is more settled it
will have a lower surface and will not obscure the outer wall as much, and so
the inner disk could extend out to further radii.  
Also, our simulations do not include optically thin dust within the gap and 
we note that we did not include any possible shadowing of dust in the optically
thin regions inside the gaps of LkCa~15 and Rox~44.
Infrared interferometry is crucial to
explore how the spatial distribution of this component is shaped as it
filters into the gap from the outer disk \citep{rice06} or how it is
influenced by possible planets in the gap \citep{lubow06}.

\subsection{Dust Opacities} \label{sec:cav}

The opacity of the disk is controlled by dust and in any sophisticated
disk model the largest uncertainty lies in the adopted dust opacities.
The dust opacity sets the location of the disk surface, i.e.~where the
optical depth to the stellar radiation reaches $\sim 1$, and in turn the
amount of disk flaring. The shape of the disk surface dictates the
fraction of the stellar radiation that is captured by the disk, and as a
result, the heating of the disk and the ensuing disk emission
\citep{calvet91}.

\begin{figure}
\epsscale{1.0}
\plotone{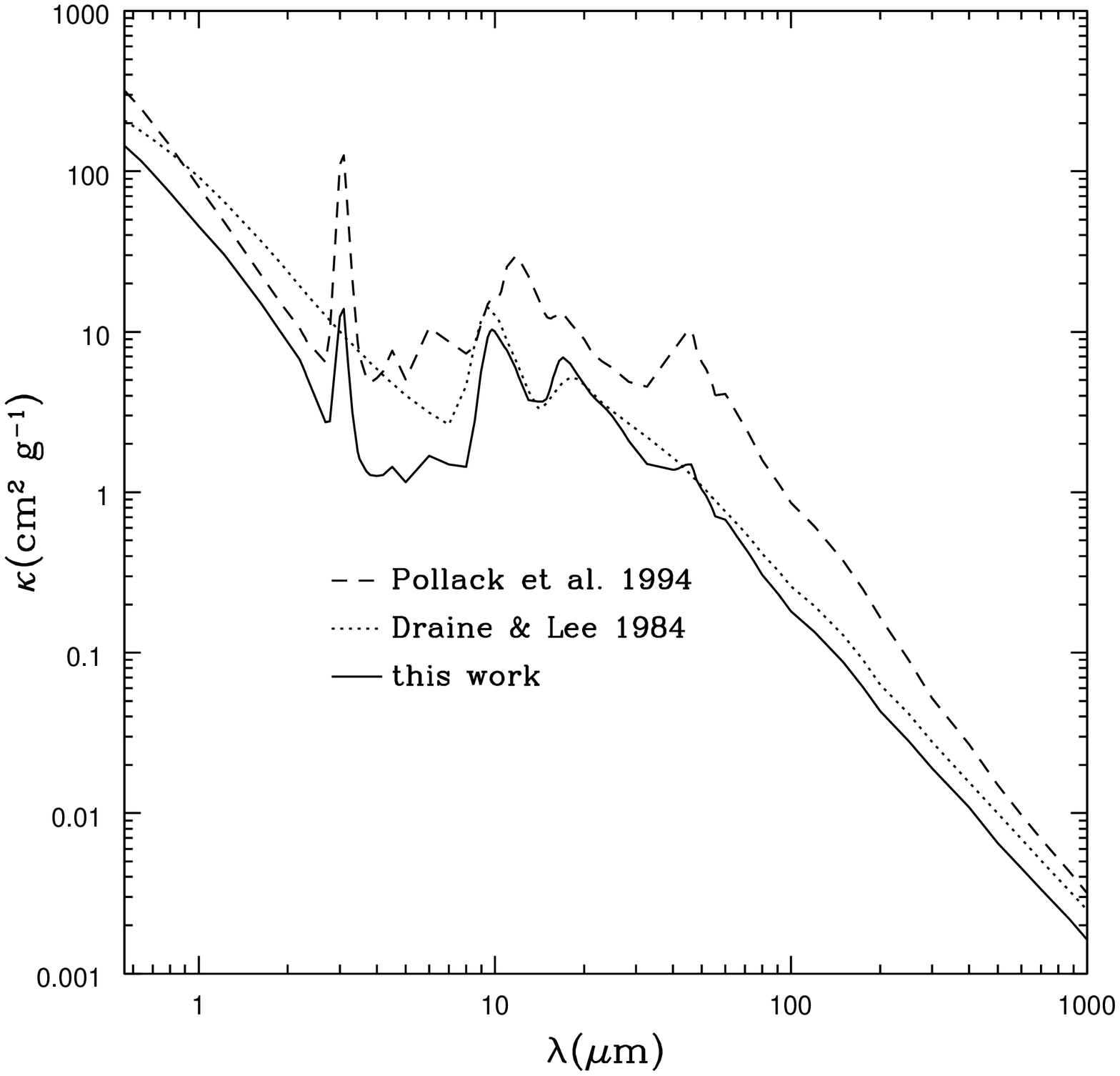}
\caption[Opacity.] {Comparison of the opacities adopted in this work to those
from \citet{pollack94} and \citet{draine84}.  
}
\label{opacity}
\end{figure}

The SED models derived in this work are obviously tied to the dust
opacity that is assumed in {\S}~\ref{sec:mod}.  In particular,
R$_{wall}^o$ depends on the adopted opacity following
Eq.~\ref{rdust}. We find that by adopting the dust opacities of
\citet{draine84}, the derived R$_{wall}^o$ of our targets increases by
$\sim$10~AU. This occurs because the opacity to the stellar radiation
($\kappa_s$), which peaks around 1~{\micron}, is higher for the dust
composition of \citet{draine84} than that adopted in this work
(Figure~\ref{opacity}). For \citet{pollack94}'s opacities, R$_{wall}^o$
decreases by $\sim$10~AU. In this case, the opacity to the disk's local
emission ($\kappa_d$), which peaks $\sim$25~{\micron}, is much higher
than ours.

The major differences between the opacities we adopt here and those of
\citet{draine84} and \citet{pollack94} are the assumed carbon component
and the ice abundance.  While most models assume some carbon component
in their dust composition, there is yet no general consensus on the
species of the carbon grains. \citet{draine84} assume that the carbon
component is due to graphite and \citet{pollack94} assume it is due to
organics. 
Further study is needed to better determine the carbon component
in accretion disks.
The {\it Herschel} mission is
an excellent opportunity to learn more about water ice in disks.
\citet{pollack94} adopt a dust-to-gas mass ratio for water ice of 0.0056
while \citet{dalessio06} find that the features produced by this value
do not agree with observations of disks. {\it Herschel} can help settle
this issue by detecting water lines in the far-infrared and allowing
derivations of the abundance of water in disks.

\subsection{Implications of Pre-Transitional Disks on Dust Clearing
Mechanisms}

Several mechanisms have been proposed to explain disk clearing.
These include planets, grain growth, the magnetorotational instability,
and photoevaporation by the central star.
The structure of pre-transitional disks - an inner disk, a gap, and an outer disk -
is predicted by planet formation models.
On the other hand, the three alternative clearing mechanisms struggle to make 
a gapped disk structure and, due to 
their predominantly inside-out clearing pattern, are instead more likely to form a central hole in a disk. 

Given the simple stipulation of a disk gap, planets emerge as the most likely dust
clearing mechanism in pre-transitional disks. 
To further explore this possibility, one must ascertain if
planet formation timescales are in line with the ages of
these disks,  if there is enough mass in these disks to have formed planets, and if
planets can open the kind of gaps that we are detecting.
Assuming that planets are indeed present within these disk gaps, one can attempt to place
some constraints on theoretical planet formation models.

It is feasible to form planets within 1 Myr, the age of our pre-transitional disk sample,
via the two leading theories of planet
formation: the gas instability model and the core nucleated accretion model. 
In the gas instability model,
fragmentation due to gravitational instabilities in the dense dust
midplane \citep{goldreich73} can form Jupiter-mass clumps in the outer
disk within a few hundred years \citep{boss00, mayer02, durisen07}.
According to the core nucleated accretion model \citep{pollack96,
bodenheimer00}, dust grains will grow into planetesimals which will
accrete other planetesimals and then form solid cores.  These cores can
become terrestrial planets or, if the cores are massive enough, gas
giant cores which can accrete the surrounding gas in the disk
\citep{pollack96,bodenheimer00}.
For several years it was thought that the formation time for
giant planets via core accretion was longer than the lifetime of
disks, however, more recent simulations show that giant planets
can form in the inner disk within $\sim$1~Myr via two different mechanisms. 
One possible mechanism which could
accelerate planet formation is migration.  
\citet{alibert05} found that
a migrating core will not deplete its surrounding feeding zone and will
reach its cross-over mass, the point at which the planet enters runaway
gas accretion,  sooner than found in previous models. 
Alternatively, the
time to enter runaway gas accretion can be reduced by adopting low dust
opacities \citep{pollack96}. 
A Jupiter-mass
planet with a core of 10 Earth-masses can form at 5~AU in 1~Myr 
\citep{hubickyj05}, about half the time predicted by models using higher
dust opacities.

It is likely that the pre-transitional disks originally had enough mass
within their presently cleared regions to have formed multiple giant planets {\it in situ} and that planet
formation may still be underway. Compared to a full disk with the same
model parameters (Table~\ref{tab:prop}), LkCa~15, UX~Tau~A, and Rox~44
are ``missing'' 10~M$_{Jupiter}$, 40~M$_{Jupiter}$, and
20~M$_{Jupiter}$, respectively, some fraction of which could have gone
into making giant planets. There is still enough material in the
outer disk of these objects (Table~\ref{tab:prop}) to potentially go
into forming planets. 

The next issue to address concerns the nature of the signatures that
forming planets will leave behind in young disks. Theories predict that
newly forming planets should clear the material around themselves
through tidal disturbances leaving behind disk gaps \citep{goldreich80, ward88, rice03,
paardekooper04, quillen04, varniere06}. 
To clear out the relatively large gaps we measure,
it is likely that more than one planet is present and
detailed simulations on the effects of multiple planets on the dust distribution
are necessary.
It is already known that when a stellar companion is located in the 
inner disk, the dust is dynamically cleared
\citep{mathieu91, artymowicz94}.
One example is the cavity in the inner
disk of CoKu Tau$/4$ which is most likely due to clearing by its stellar
companion \citep{ireland08} and whose SED has been reproduced by a
circumbinary disk model \citep{nagel10}. 
The object T54 in the Chamaeleon cloud is also a likely candidate for dust
clearing by a stellar companion \citep{kim09}.  
These observations are evidence that dynamical dust clearing by companions
does occur; the salient question is whether the companion in pre-transitional disks
is in the stellar-mass or planet-mass regime.
Searches for companions in the
disks of UX~Tau~A and LkCa~15 have revealed that they are single stars
down to 0.35~AU \citep[][A. Kraus, private communication]{pott10,
white01}.  To explain the location of the outer walls of UX Tau A and
LkCa~15, a stellar companion would have to be located at larger radii
\citep{artymowicz94}, therefore we can conclude that the truncation of
the outer disk is not due to a stellar companion. Multiplicity studies
of Rox~44 down to similar small radii do not yet exist.

If planets are clearing the dust in pre-transitional disks, we can
speculate on the possible masses of such planets and their most likely
formation scenario. Simulations of disk clearing demonstrate that
Neptune-mass planets ($\sim$0.05~M$_{Jupiter}$) can open gaps in the
dust component of the disk while planets with 0.5~M$_{Jupiter}$ can
create a gap in the gas disk \citep{paardekooper06}. We can conjecture
that the planets in the pre-transitional disks are between 0.05
M$_{Jupiter}$--0.5 M$_{Jupiter}$ based on the detection of a gap in
their dust disks and evidence which suggests there are not substantial
gaps in the gas disk. Since these disks are actively accreting, we can
assume that gas travels from the outer disk to the inner disk and that
therefore gas is located within the dust gap. In fact, in LkCa~15, UX~Tau~A, and
Rox~44, gas has been detected within the dust gap
\citep{najita03,salyk09}. These partially-filled gas gaps have
consequences on Type II migration. According to classical Type II
migration, planets will migrate inward onto the star on the viscous
timescale, followed by the outer disk edge, and therefore we would not
see gaps on these short timescales \citep{lin86}. The gap in the gas
disk in the above scenario is deep and there is no gas flow across it.
However, \citet{crida07} showed that inward planet migration can be
slowed or stopped if the 
gap is not completely cleared of gas. In this
case, the gas in the gap will exert a positive torque on the outer disk
which diminishes the total torque that the planet feels from the outer
disk.  As a result the outer disk will not migrate the planet inward as
quickly, opening the possibility that planets can sustain gaps in disks
long enough for us to detect them. Based on the points above, 
it follows that the planets in pre-transitional disks were 
most likely formed via core
accretion since gravitational instability tends to form planets of
several Jupiter masses in the outermost disk, and moreover, 
these planets would have needed to
migrate to the inner disk to form the gaps we are detecting.

The alternative disk clearing mechanisms proposed to date have
difficulties explaining several observations and are most likely not responsible for
the type of clearing seen in pre-transitional disks.
Grain
growth simulations predict that dust will evolve on faster timescales in
the inner parts of the disk, and eventually the small grains which
contribute to the near-IR emission of the disk will grow causing a flux
deficit in the SED \citep{dullemond04,dullemond05}. 
However, the gap edges and silicate features are too sharp to be explained by grain growth.
In addition, grain growth is an inside-out clearing mechanism and does not 
account for a remnant inner disk.

In the MRI clearing scenario,
 X-rays from the star can activate the MRI in the ionized
inner wall of the disk which will lead material to accrete from the wall
onto the star, creating a hole in the disk that grows outward
\citep{chiang07}.  It predicts a stronger trend of gap size with X-ray luminosity 
and accretion rate than is observed \citep{kim09} and requires an
inner disk hole, not a gap, to already exist before it can take effect.

Radiation from the central star can potentially photoevaporate the
surrounding disk \citep{hollenbach94,clarke01}. 
High energy photons from
the stellar wind will impinge upon the upper disk layers.  As the disk evolves
viscously, the mass accretion rate decreases with time, eventually
reaching the mass loss rate in the photoevaporative wind. At this point,
the photoevaporative wind takes over and inward accretion onto the star will
essentially stop.  Once the isolated inner disk drains onto the star on
the viscous timescale, the inner edge of the outer disk will be directly
irradiated by the star and the hole will keep growing outward \citep{alexander07}. 
The typical value
of the mass loss rate in the photoevaporative wind was typically taken
to be $\sim$4$\times$10$^{-10}$~{\msun}~yr$^{-1}$ based on EUV radiation
alone \citep{clarke01}, however, this value has recently jumped to
$\sim$10$^{-8}$~{\msun}~yr$^{-1}$ due to the inclusion of X-ray and UV
photons \citep{gorti09a, owen10}. 
However, if mass loss rates are as high as this value it poses significant
problems.  The average mass accretion rate
measured in TTS is $\sim$10$^{-8}$~{\msun}~yr$^{-1}$ \citep{hartmann98}.
 According to the evolutionary scenario outlined above, it follows that we should not see disks with
mass accretion rates lower than this value, in contrast to what is
observed \citep{hartmann98}. Moreover, modeling of disk wind
indicators (i.e. forbidden line profiles) by \citet{hartigan95} find
that the winds of TTS are typically $\sim$0.01 of the mass accretion
rate.  In any case, 
photoevaporation opens short-lived gaps in the disk that would not lead
to a discernible deficit in the  
SED (see  $\S$~\ref{sec:ptd}), and the models presented to date are not
compatible with large 40 -- 50~AU gaps in massive disks which are still actively
accreting. Furthermore, due to shadowing, there are 
portions of the outer wall that will not be affected by the
photoevaporative wind. 

Given the detection of over one hundred
Jupiter-mass planets \citep{butler06} and about one dozen Neptune-mass
planets (see references within \citet{endl08}), it is natural to wonder if we can see
evidence for the incipient stages of planet formation and particularly
if pre-transitional disks around TTS constitute such evidence.
At the moment, {\it ALMA} is the best suited facility to
detect disks around young planets within TTS disks \citep{wolf05}
and therefore search for planets within the gaps of pre-transitional disks. 
{\it ALMA}, along with {\it JWST}, will help us to better constrain the statistics
of pre-transitional disks and their evolution.  
High-resolution near-IR interferometry can also examine the inner disk
and reveal the spatial distribution of the optically thin dust within
the disk gap.

\section{Summary \& Conclusions} \label{sed:sum}

In this paper we presented spectral type measurements as well as veiling
and near-infrared excess measurements for the diskless LkCa~14, the
pre-transitional disks of LkCa~15, UX~Tau~A, and Rox~44, and the
transitional disks of DM~Tau and GM~Aur. Using near-IR SpeX spectra from
1--5 {\micron}, we extracted the near-IR excess continua of our
pre-transitional disk sample and fit their excess emission with
optically thick inner disk walls, supporting the interpretation that
these objects contain gaps in their disks. We modeled the broad-band
SEDs of these pre-transitional disks and measure gap sizes of
$\sim$40--70~AU.  In the case of LkCa~15 and Rox~44 there is
$\sim$10$^{-11}$~{\msun} of ISM-sized optically thin dust within the
gap.

The near-infrared emission of our pre-transitional disks differs
significantly from that of the transitional disks of DM~Tau and GM~Aur.
DM~Tau has no veiling and no excess emission in the near-IR, indicating
that its disk hole is relatively empty of small dust grains.  GM~Aur has
a small amount of veiling and displays a flat excess above the stellar
photosphere in the near-IR, which can be fit by a model of optically
thin dust emission.  This is in contrast to the pre-transitional disks
which have large, blackbody-like near-IR excess continua that can be fit
with models of a wall located at the dust sublimation radius.

We also studied the effects of shadowing of the outer disk wall
by the inner disk. We found that when the finite size of the star
is taken into account, a significant portion of the outer wall
is either in the prenumbra or completely out of the shadow of the inner
wall. The predicted height of the wall is consistent with
that of the outer disk, which is derived independently.

Based on currently known disk clearing mechanisms, we propose that the
gaps in pre-transitional disks are indicators of planet formation,
making these disks a promising location for young planet searches. {\it
ALMA} will be pivotal in extending the sample of known pre-transitional
disks and has the potential to detect planets in these gaps. Near-IR
interferometry will play an important role in further understanding the
innermost regions of these disks. Future studies of this class of
objects may bring us a few steps closer to understanding the origin of
our own solar system.

 \acknowledgments{We thank Cesar Brice{\~n}o and Scott Kenyon for kindly
 providing the FAST spectra used in this paper. We also thank Perry
 Berlind for performing the FAST observations.  We thank Sean Andrews
 and David Wilner for useful discussions. The near-infrared spectra used in
 this paper were obtained using SpeX at the Infrared Telescope Facility,
 which is operated by the University of Hawaii under Cooperative
 Agreement no.~NNX08AE38A with the National Aeronautics and Space
 Administration, Science Mission Directorate, Planetary Astronomy
 Program. C.~E. was supported by the National Science Foundation under
 Award No. 0901947.  
 P.~D. acknowledges a grant from PAPIIT-DGAPA UNAM.
 E.~N. thanks a postdoctoral fellowship from Conacyt.
 K.~L. was supported by grant AST-0544588 from the National
 Science Foundation.  N.~C. acknowledges support from NASA Origins Grant
 NNX08AH94G. }

\appendix
\section{The Effect of Disk Shadowing} \label{sec:sha}

A point source does not cast a penumbra.  It only casts an umbra with a size, 
measured from above the
midplane, of \be z_{point}= z_{w} {r \over r_w} \en where $z_{w}$ is the
height of the inner wall above the midplane, $r_w$ is the radial
distance between the center of the star and the inner wall, and $r$ is
the radius of the outer wall as measured from the center of the star.

\begin{figure}
\epsscale{1}
\plotone{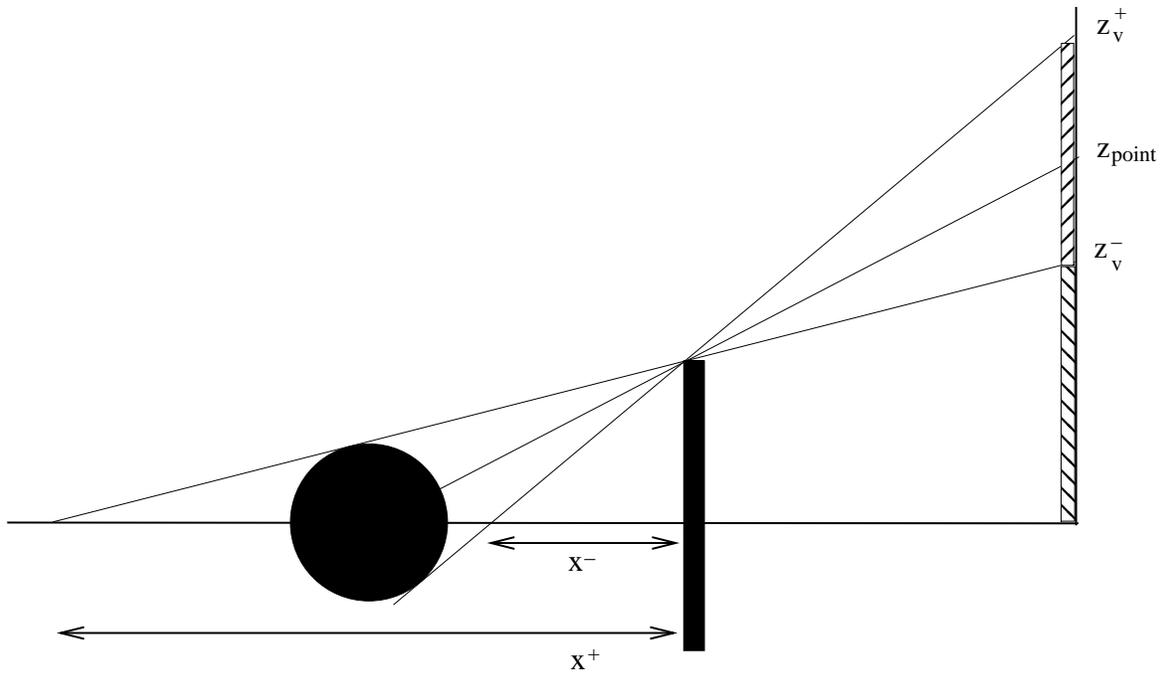}
\caption[Limits of the shadow cast by a finite source.] {Limits of the
shadow cast by a finite source.  Above $z_v^+$ the whole star is visible
and below $z_v^-$ the star is in the umbra. In the intermediate region,
the surface of the star is partially visible (i.e. the wall is in the
penumbra). $z_{point}$ is the height
of the umbra when the star is taken to be a point source.}
\label{shadow1}
\end{figure}

However, since the star is a finite source, the size of the umbra will
be much smaller. At the outer wall, the height over which the whole star
can be seen and the height below which the star cannot be seen are
calculated using the roots of a quadratic function. This function is
constructed for the distance between the inner wall and the points where
rays tangent to the upper and lower parts of the star and the upper part
of the inner wall cross the horizontal plane (Figure~\ref{shadow1}).
Using dimensionless distances (with  $R_*$ as unity)
\be
x^+ = (r_w + \sqrt{(1 + {1 \over z_w})(1- {1 \over z_w})+{ r_w^2 \over z_w^2}})/ 
[(1 + {1 \over z_w})(1- {1 \over z_w})],
\en
\be
x^- = (r_w - \sqrt{(1 + {1 \over z_w})(1- {1 \over z_w})+{ r_w^2 \over z_w^2}})/ 
[(1 + {1 \over z_w})(1- {1 \over z_w})],
\en
\be
z_v^+=(r-r_w)(z_w/x^-)+z_w\label{eqn:zp},
\en
\be
z_v^-=(r-r_w)(z_w/x^-)+z_w\label{eqn:zn},
\en
and, consequently, the outer wall above $z_v^-$ will be illuminated to some
degree.

Given that a significant portion of the outer wall is not fully
illuminated by the star, we consider the effect of the shadow,
both the umbra and penumbra, on the heating of the wall and hence the
resulting emission.  The solid angle subtended by the star as seen by
any point  $P$ in the disk, with coordinates $r$ and $z$, is
\be
\Omega_* (r,z) = 2 \int_0^{\theta_{max}} \int_0^{\phi_{max}} 
 sin \phi d\phi d\theta  \label{eqn:omega}
\en
where $\phi$ characterizes an annulus at the stellar surface, and
$\theta$ allows one to integrate over the annulus. The maximum value of
$\phi$ is
\be
\phi_{max}= sin^{-1}(1/d)
\en
where 
\be
d=(z^2+r^2)^{1/2}.
\en
The maximum value of $\theta$ depends on the shadow.

\begin{figure}
\epsscale{1}
\plotone{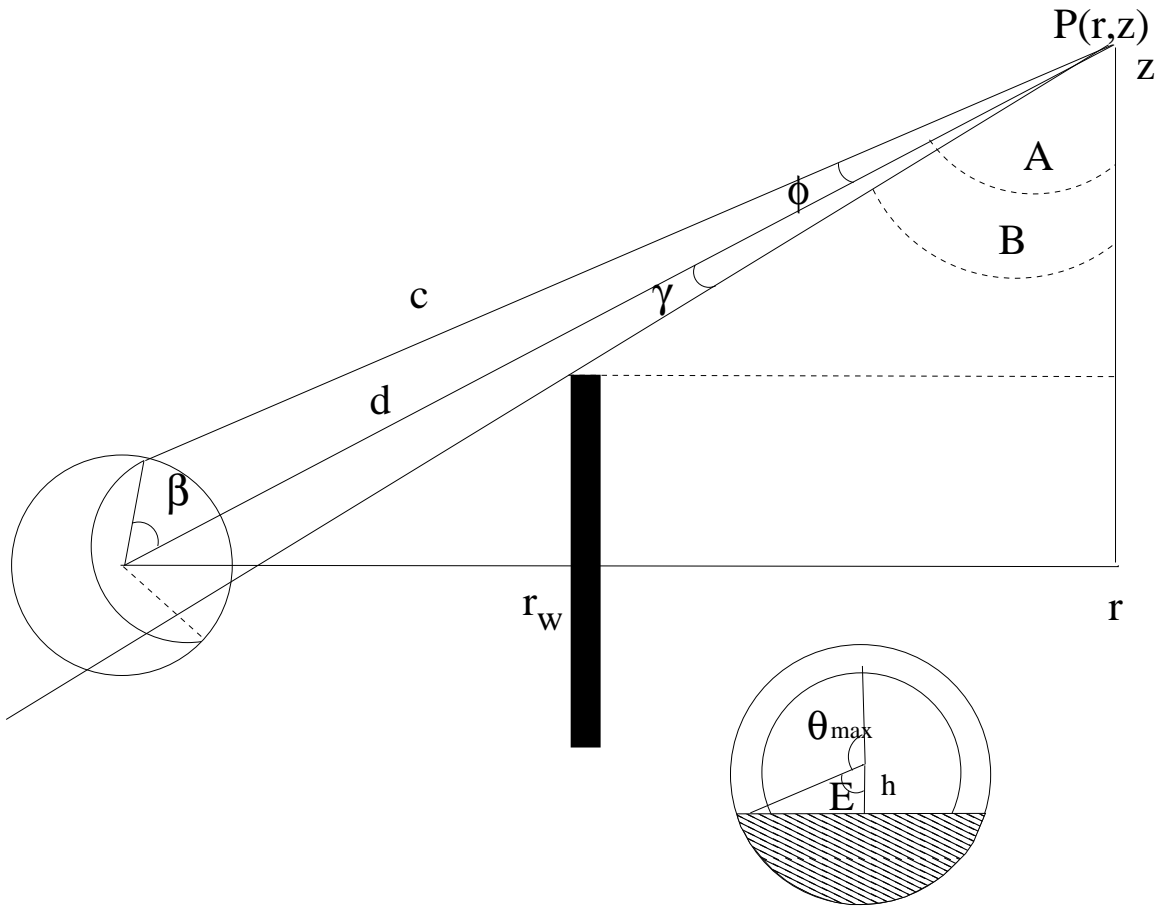}
\caption[View of observer at outer wall.] {Geometry of the star, inner
wall, and outer wall used to define the integration limits of the solid
angle subtended by the star.}
\label{shadow2}
\end{figure}

In Figure~\ref{shadow2} we show the relevant angles used to define
the integration limits of the solid angle of the star. The angle $\gamma$
is related to the height of the shadow. If $\phi < \gamma$ the
whole annulus can be seen, and then $\theta_{max}=\pi$. However, if
$\phi > \gamma$ then a fraction of the annulus is in the shadow.

From the figure, we can see that 
\be
\theta_{max} = \pi -E \label{eqn:theta}
\en
and 
\be
h=\sin \beta \cos E.
\en
To find the angle $\beta$ we use that
\be
\sin \beta = c \sin \phi
\en
and 
\be
c= d \cos \phi - \sqrt{1-(d \sin \phi)^2}.
\en
Then,  $\beta$ is given by
\be
\sin \beta= d \cos \phi \sin \phi - 
[1-(d \sin \phi)^2]^{1/2}  \sin \phi.
\en
The other relevant angle $\gamma$, is given by $\gamma=A-B$, and so 
\be
\gamma= \tan^{-1} (r/z) - \tan^{-1} [(r-r_w)/(z-z_w)].
\en
The height $h$ is given by
\be
h=(d-\cos \beta) \tan \gamma
\en
and thus, 
\be
\cos E = (d - \cos \beta) \tan \gamma/\sin \beta,
\en
which is used in Eq.~\ref{eqn:theta} to calculate $\theta_{max}$,
which in turn allows one to calculate the solid angle, $\Omega_*$,  with Eq.~\ref{eqn:omega}.

\clearpage

\begin{deluxetable}{lcccc}
\tabletypesize{\scriptsize}
\tablewidth{0pt}
\tablecaption{Log of SpeX Observations \label{tab:spexlog}}
\startdata
\hline
Target & K$_{s}^{1}$  & Date & LXD Exposure (s) & SNR$^{2}$    \\
\hline
LkCa~14      &  8.6 & 2009 Feb 18  &  1800 & 270\\
LkCa~15      & 8.2 & 2009 Dec 22   & 2400  & 370\\
UX~Tau~A   & 7.6 & 2009 Dec 22   & 2400  & 450\\
Rox~44        & 7.6 & 2009 Apr 07   & 1800  & 150\\
GM~Aur       & 8.3 &  2009 Dec 22  &   2400 & 300\\
DM~Tau      &  9.5 & 2009 Feb 16  &  2400 & 170

\enddata
\tablenotetext{1}{Two-Micron All-Sky Survey Point Source Catalog \citep[2MASS;][]{skrutskie06}}
\tablenotetext{2}{Measured at $\sim$2.2~{\micron}.}
\end{deluxetable}

\begin{deluxetable}{lcccc}
\tabletypesize{\scriptsize}
\tablewidth{0pt}
\tablecaption{Characteristics of Sample\label{tab:spexchar}}
\startdata
\hline
Target & Spectral & r$_K$ & T$_{wall}^i$ & Disk\\
  & Type & &  (K)$^{1}$ & Class$^{2}$ \\
\hline
LkCa~14     &  K5.5  &      0  & ... & diskless\\ 
LkCa~15     &  K3  & 0.6 &1400 & pTD\\  
UX~Tau~A   &  G8 &   0.4 & 1550 & pTD\\ 
Rox~44       & K3$^{3}$ &  0.8 & 1200 & pTD\\  
GM~Aur       &  K5.5  &   0.1 & ... & TD\\  
DM~Tau      &  M1.5  &  0 & ... & TD

\enddata
\tablenotetext{1}{T$_{wall}^i$ refers to the temperature of the best-fit
wall to the the excess near-IR emission, if present.}
\tablenotetext{2}{Our targets are classified as diskless stars, stars
with transitional disks (TD), and stars with pre-transitional disks
(pTD).}
\tablenotetext{3}{Value adopted from \citet{bouvier92}.}
\end{deluxetable}

\begin{deluxetable}{l c c c}
\tabletypesize{\scriptsize}
\tablewidth{0pt}
\tablecaption{Stellar and Model Properties\\ of LkCa~15, UX~Tau~A,  \& Rox~44\label{tab:prop}}
\startdata
\hline
\hline
\multicolumn{4}{c}{Stellar Properties}\\
\hline
\hline
\colhead{} & \colhead{LkCa~15}  & \colhead{UX~Tau~A} & \colhead{Rox~44}\\
\hline
M$_{*}$ (M$_{\sun}$) & 1.3  & 1.5 & 1.3\\
R$_{*}$ (R$_{\sun}$) & 1.6  & 1.8 & 1.6\\
T$_{*}$ (K) & 4730  & 5520 & 4730 \\
$\mdot$ (M$_{\sun}$ yr$^{-1}$) & 3.3$\times$10$^{-9}$ & 1.1$\times$10$^{-8}$ & 9.3$\times$10$^{-9}$\\
Inclination (deg) & 42$^{1}$ & 60 & 45$^{2}$\\
Distance (pc) & 140& 140 & 120\\
A$_V$ & 1.7 & 1.8 & 2.2\\
\hline
\hline
\multicolumn{4}{c}{Optically Thick Inner Wall}\\
\hline
a$_{max}$ ({\micron}) & 1 & 10 & 1\\
T$_{wall}^i$ (K) & 1400  & 1550 & 1200\\
z$_{wall}^i$ (AU) & 0.017  & 0.009& 0.034\\
R$_{wall}^i$ (AU) & 0.15 & 0.15  & 0.25\\
\hline
\multicolumn{4}{c}{Optically Thick Inner Disk}\\
\hline
R$_{inner~disk}$ (AU) & $<$0.19  & $<$0.21 & $<$0.4\\
M$_{inner~disk}$ (M$_{\sun}$) & $<$2$\times$10$^{-4}$ & $<$6$\times$10$^{-5}$ & $<$8$\times$10$^{-5}$\\
\hline
\multicolumn{4}{c}{Optically Thick Outer Wall} \\
\hline
a$_{max}$ ({\micron}) & 0.25  & 0.25 & 0.25\\
T$_{wall}^o$ (K)$^3$ & 95  & 110 & 120\\
z$_{wall}^o$ (AU) & 12.9 & 13.8 & 9.9\\
R$_{wall}^o$ (AU)$^3$ & 58 & 71 & 36\\
\hline
\multicolumn{4}{c}{Optically Thick Outer Disk}\\
\hline
$\epsilon$ & 0.001  & 0.001 & 0.01\\
$\alpha$ & 0.0005  & 0.004 & 0.006\\
M$_{disk}$ (M$_{\sun}$) & 0.1 & 0.04 &~0.03
\enddata
\tablenotetext{1}{\citet{simon00}}
\tablenotetext{2}{\citet{andrews09}}
\tablenotetext{3}{Since we take into account the varying illumination along the outer wall due to the shadowing,
there will be a range of T$_{wall}^o$.  Here we report the temperature of the fully illuminated part 
of the wall and use this value to calculate R$_{wall}^o$.}
\end{deluxetable}

\clearpage

\begin{thebibliography}{109}
\expandafter\ifx\csname natexlab\endcsname\relax\def\natexlab#1{#1}\fi

\bibitem[{{Alexander} \& {Armitage}(2007)}]{alexander07}
{Alexander}, R.~D., \& {Armitage}, P.~J. 2007, \mnras, 375, 500

\bibitem[{{Alibert} {et~al.}(2005){Alibert}, {Mordasini}, {Benz}, \&
  {Winisdoerffer}}]{alibert05}
{Alibert}, Y., {Mordasini}, C., {Benz}, W., \& {Winisdoerffer}, C. 2005, \aap,
  434, 343

\bibitem[{{Andrews} \& {Williams}(2005)}]{andrews05}
{Andrews}, S.~M., \& {Williams}, J.~P. 2005, \apj, 631, 1134

\bibitem[{{Andrews} {et~al.}(2009){Andrews}, {Wilner}, {Hughes}, {Qi}, \&
  {Dullemond}}]{andrews09}
{Andrews}, S.~M., {Wilner}, D.~J., {Hughes}, A.~M., {Qi}, C., \& {Dullemond},
  C.~P. 2009, \apj, 700, 1502

\bibitem[{{Artymowicz} \& {Lubow}(1994)}]{artymowicz94}
{Artymowicz}, P., \& {Lubow}, S.~H. 1994, \apj, 421, 651

\bibitem[{{Begemann} {et~al.}(1994){Begemann}, {Dorschner}, {Henning},
  {Mutschke}, \& {Thamm}}]{begemann94}
{Begemann}, B., {Dorschner}, J., {Henning}, T., {Mutschke}, H., \& {Thamm}, E.
  1994, \apjl, 423, L71

\bibitem[{{Bergin} {et~al.}(2004){Bergin}, {Calvet}, {Sitko}, {Abgrall},
  {D'Alessio}, {Herczeg}, {Roueff}, {Qi}, {Lynch}, {Russell}, {Brafford}, \&
  {Perry}}]{bergin04}
{Bergin}, E., {et~al.} 2004, \apjl, 614, L133

\bibitem[{{Bodenheimer} {et~al.}(2000){Bodenheimer}, {Hubickyj}, \&
  {Lissauer}}]{bodenheimer00}
{Bodenheimer}, P., {Hubickyj}, O., \& {Lissauer}, J.~J. 2000, Icarus, 143, 2

\bibitem[{{Boss}(2000)}]{boss00}
{Boss}, A.~P. 2000, \apjl, 536, L101

\bibitem[{{Bouvier} \& {Appenzeller}(1992)}]{bouvier92}
{Bouvier}, J., \& {Appenzeller}, I. 1992, \aaps, 92, 481

\bibitem[{{Butler} {et~al.}(2006){Butler}, {Wright}, {Marcy}, {Fischer},
  {Vogt}, {Tinney}, {Jones}, {Carter}, {Johnson}, {McCarthy}, \&
  {Penny}}]{butler06}
{Butler}, R.~P., {et~al.} 2006, \apj, 646, 505

\bibitem[{{Calvet} {et~al.}(2002){Calvet}, {D'Alessio}, {Hartmann}, {Wilner},
  {Walsh}, \& {Sitko}}]{calvet02}
{Calvet}, N., {D'Alessio}, P., {Hartmann}, L., {Wilner}, D., {Walsh}, A., \&
  {Sitko}, M. 2002, \apj, 568, 1008

\bibitem[{{Calvet} \& {Gullbring}(1998)}]{calvet98}
{Calvet}, N., \& {Gullbring}, E. 1998, \apj, 509, 802

\bibitem[{{Calvet} {et~al.}(1992){Calvet}, {Magris}, {Patino}, \&
  {D'Alessio}}]{calvet92}
{Calvet}, N., {Magris}, G.~C., {Patino}, A., \& {D'Alessio}, P. 1992, Revista
  Mexicana de Astronomia y Astrofisica, 24, 27

\bibitem[{{Calvet} {et~al.}(1991){Calvet}, {Patino}, {Magris}, \&
  {D'Alessio}}]{calvet91}
{Calvet}, N., {Patino}, A., {Magris}, G.~C., \& {D'Alessio}, P. 1991, \apj,
  380, 617

\bibitem[{{Calvet} {et~al.}(2005){Calvet}, {D'Alessio}, {Watson},
  {Franco-Hern{\'a}ndez}, {Furlan}, {Green}, {Sutter}, {Forrest}, {Hartmann},
  {Uchida}, {Keller}, {Sargent}, {Najita}, {Herter}, {Barry}, \&
  {Hall}}]{calvet05}
{Calvet}, N., {et~al.} 2005, \apjl, 630, L185

\bibitem[{{Chiang} \& {Murray-Clay}(2007)}]{chiang07}
{Chiang}, E., \& {Murray-Clay}, R. 2007, Nature Physics, 3, 604

\bibitem[{{Clarke} {et~al.}(2001){Clarke}, {Gendrin}, \&
  {Sotomayor}}]{clarke01}
{Clarke}, C.~J., {Gendrin}, A., \& {Sotomayor}, M. 2001, \mnras, 328, 485

\bibitem[{{Cohen} \& {Kuhi}(1979)}]{cohen79}
{Cohen}, M., \& {Kuhi}, L.~V. 1979, \apjs, 41, 743

\bibitem[{{Crida} \& {Morbidelli}(2007)}]{crida07}
{Crida}, A., \& {Morbidelli}, A. 2007, \mnras, 377, 1324

\bibitem[{{Cushing} {et~al.}(2005){Cushing}, {Rayner}, \& {Vacca}}]{cushing05}
{Cushing}, M.~C., {Rayner}, J.~T., \& {Vacca}, W.~D. 2005, \apj, 623, 1115

\bibitem[{{Cushing} {et~al.}(2004){Cushing}, {Vacca}, \& {Rayner}}]{cushing04}
{Cushing}, M.~C., {Vacca}, W.~D., \& {Rayner}, J.~T. 2004, \pasp, 116, 362

\bibitem[{{D'Alessio} {et~al.}(2006){D'Alessio}, {Calvet}, {Hartmann},
  {Franco-Hern{\'a}ndez}, \& {Serv{\'{\i}}n}}]{dalessio06}
{D'Alessio}, P., {Calvet}, N., {Hartmann}, L., {Franco-Hern{\'a}ndez}, R., \&
  {Serv{\'{\i}}n}, H. 2006, \apj, 638, 314

\bibitem[{{D'Alessio} {et~al.}(1999){D'Alessio}, {Calvet}, {Hartmann},
  {Lizano}, \& {Cant{\'o}}}]{dalessio99}
{D'Alessio}, P., {Calvet}, N., {Hartmann}, L., {Lizano}, S., \& {Cant{\'o}}, J.
  1999, \apj, 527, 893

\bibitem[{{D'Alessio} {et~al.}(2005){D'Alessio}, {Hartmann}, {Calvet},
  {Franco-Hern{\'a}ndez}, {Forrest}, {Sargent}, {Furlan}, {Uchida}, {Green},
  {Watson}, {Chen}, {Kemper}, {Sloan}, \& {Najita}}]{dalessio05}
{D'Alessio}, P., {et~al.} 2005, \apj, 621, 461

\bibitem[{{Dorschner} {et~al.}(1995){Dorschner}, {Begemann}, {Henning},
  {Jaeger}, \& {Mutschke}}]{dorschner95}
{Dorschner}, J., {Begemann}, B., {Henning}, T., {Jaeger}, C., \& {Mutschke}, H.
  1995, \aap, 300, 503

\bibitem[{{Draine} \& {Lee}(1984)}]{draine84}
{Draine}, B.~T., \& {Lee}, H.~M. 1984, \apj, 285, 89

\bibitem[{{Dullemond} \& {Dominik}(2004)}]{dullemond04}
{Dullemond}, C.~P., \& {Dominik}, C. 2004, \aap, 421, 1075

\bibitem[{{Dullemond} \& {Dominik}(2005)}]{dullemond05}
---. 2005, \aap, 434, 971

\bibitem[{{Durisen} {et~al.}(2007){Durisen}, {Boss}, {Mayer}, {Nelson},
  {Quinn}, \& {Rice}}]{durisen07}
{Durisen}, R.~H., {Boss}, A.~P., {Mayer}, L., {Nelson}, A.~F., {Quinn}, T., \&
  {Rice}, W.~K.~M. 2007, in Protostars and Planets V, ed. B.~{Reipurth},
  D.~{Jewitt}, \& K.~{Keil}, 607--622

\bibitem[{{Edwards} {et~al.}(2006){Edwards}, {Fischer}, {Hillenbrand}, \&
  {Kwan}}]{edwards06}
{Edwards}, S., {Fischer}, W., {Hillenbrand}, L., \& {Kwan}, J. 2006, \apj, 646,
  319

\bibitem[{{Endl} {et~al.}(2008){Endl}, {Cochran}, {Wittenmyer}, \&
  {Boss}}]{endl08}
{Endl}, M., {Cochran}, W.~D., {Wittenmyer}, R.~A., \& {Boss}, A.~P. 2008, \apj,
  673, 1165

\bibitem[{{Espaillat}(2009)}]{espaillat09}
{Espaillat}, C. 2009, PhD thesis, University of Michigan

\bibitem[{{Espaillat} {et~al.}(2007{\natexlab{a}}){Espaillat}, {Calvet},
  {D'Alessio}, {Hern{\'a}ndez}, {Qi}, {Hartmann}, {Furlan}, \&
  {Watson}}]{espaillat07b}
{Espaillat}, C., {Calvet}, N., {D'Alessio}, P., {Hern{\'a}ndez}, J., {Qi}, C.,
  {Hartmann}, L., {Furlan}, E., \& {Watson}, D.~M. 2007{\natexlab{a}}, \apjl,
  670, L135

\bibitem[{{Espaillat} {et~al.}(2008{\natexlab{a}}){Espaillat}, {Calvet},
  {Luhman}, {Muzerolle}, \& {D'Alessio}}]{espaillat08a}
{Espaillat}, C., {Calvet}, N., {Luhman}, K.~L., {Muzerolle}, J., \&
  {D'Alessio}, P. 2008{\natexlab{a}}, \apjl, 682, L125

\bibitem[{{Espaillat} {et~al.}(2007{\natexlab{b}}){Espaillat}, {Calvet},
  {D'Alessio}, {Bergin}, {Hartmann}, {Watson}, {Furlan}, {Najita}, {Forrest},
  {McClure}, {Sargent}, {Bohac}, \& {Harrold}}]{espaillat07a}
{Espaillat}, C., {et~al.} 2007{\natexlab{b}}, \apjl, 664, L111

\bibitem[{{Espaillat} {et~al.}(2008{\natexlab{b}}){Espaillat}, {Muzerolle},
  {Hern{\'a}ndez}, {Brice{\~n}o}, {Calvet}, {D'Alessio}, {McClure}, {Watson},
  {Hartmann}, \& {Sargent}}]{espaillat08b}
---. 2008{\natexlab{b}}, \apjl, 689, L145

\bibitem[{{Evans} {et~al.}(2009){Evans}, {Dunham}, {J{\o}rgensen}, {Enoch},
  {Mer{\'{\i}}n}, {van Dishoeck}, {Alcal{\'a}}, {Myers}, {Stapelfeldt},
  {Huard}, {Allen}, {Harvey}, {van Kempen}, {Blake}, {Koerner}, {Mundy},
  {Padgett}, \& {Sargent}}]{evans09}
{Evans}, N.~J., {et~al.} 2009, \apjs, 181, 321

\bibitem[{{Fabricant} {et~al.}(1998){Fabricant}, {Cheimets}, {Caldwell}, \&
  {Geary}}]{fabricant98}
{Fabricant}, D., {Cheimets}, P., {Caldwell}, N., \& {Geary}, J. 1998, \pasp,
  110, 79

\bibitem[{{Folha} \& {Emerson}(1999)}]{folha99}
{Folha}, D.~F.~M., \& {Emerson}, J.~P. 1999, \aap, 352, 517

\bibitem[{{Furlan} {et~al.}(2006){Furlan}, {Hartmann}, {Calvet}, {D'Alessio},
  {Franco-Hern{\'a}ndez}, {Forrest}, {Watson}, {Uchida}, {Sargent}, {Green},
  {Keller}, \& {Herter}}]{furlan06}
{Furlan}, E., {et~al.} 2006, \apjs, 165, 568

\bibitem[{{Furlan} {et~al.}(2009){Furlan}, {Watson}, {McClure}, {Manoj},
  {Espaillat}, {D'Alessio}, {Calvet}, {Kim}, {Sargent}, {Forrest}, \&
  {Hartmann}}]{furlan09}
---. 2009, \apj, 703, 1964

\bibitem[{{Goldreich} \& {Tremaine}(1980)}]{goldreich80}
{Goldreich}, P., \& {Tremaine}, S. 1980, \apj, 241, 425

\bibitem[{{Goldreich} \& {Ward}(1973)}]{goldreich73}
{Goldreich}, P., \& {Ward}, W.~R. 1973, \apj, 183, 1051

\bibitem[{{Gorti} \& {Hollenbach}(2009)}]{gorti09a}
{Gorti}, U., \& {Hollenbach}, D. 2009, \apj, 690, 1539

\bibitem[{{Gullbring} {et~al.}(1998){Gullbring}, {Hartmann}, {Briceno}, \&
  {Calvet}}]{gullbring98}
{Gullbring}, E., {Hartmann}, L., {Briceno}, C., \& {Calvet}, N. 1998, \apj,
  492, 323

\bibitem[{{Hartigan} {et~al.}(1995){Hartigan}, {Edwards}, \&
  {Ghandour}}]{hartigan95}
{Hartigan}, P., {Edwards}, S., \& {Ghandour}, L. 1995, \apj, 452, 736

\bibitem[{{Hartigan} {et~al.}(1989){Hartigan}, {Hartmann}, {Kenyon}, {Hewett},
  \& {Stauffer}}]{hartigan89}
{Hartigan}, P., {Hartmann}, L., {Kenyon}, S., {Hewett}, R., \& {Stauffer}, J.
  1989, \apjs, 70, 899

\bibitem[{{Hartigan} {et~al.}(1994){Hartigan}, {Strom}, \&
  {Strom}}]{hartigan94}
{Hartigan}, P., {Strom}, K.~M., \& {Strom}, S.~E. 1994, \apj, 427, 961

\bibitem[{{Hartmann} {et~al.}(1998){Hartmann}, {Calvet}, {Gullbring}, \&
  {D'Alessio}}]{hartmann98}
{Hartmann}, L., {Calvet}, N., {Gullbring}, E., \& {D'Alessio}, P. 1998, \apj,
  495, 385

\bibitem[{{Herbig}(1977)}]{herbig77}
{Herbig}, G.~H. 1977, \apj, 214, 747

\bibitem[{{Herbig} {et~al.}(1986){Herbig}, {Vrba}, \& {Rydgren}}]{herbig86}
{Herbig}, G.~H., {Vrba}, F.~J., \& {Rydgren}, A.~E. 1986, \aj, 91, 575

\bibitem[{{Herbst} {et~al.}(1994){Herbst}, {Herbst}, {Grossman}, \&
  {Weinstein}}]{herbst94}
{Herbst}, W., {Herbst}, D.~K., {Grossman}, E.~J., \& {Weinstein}, D. 1994, \aj,
  108, 1906

\bibitem[{{Hern{\'a}ndez} {et~al.}(2004){Hern{\'a}ndez}, {Calvet},
  {Brice{\~n}o}, {Hartmann}, \& {Berlind}}]{hernandez04}
{Hern{\'a}ndez}, J., {Calvet}, N., {Brice{\~n}o}, C., {Hartmann}, L., \&
  {Berlind}, P. 2004, \aj, 127, 1682

\bibitem[{{H{\o}g} {et~al.}(2000){H{\o}g}, {Fabricius}, {Makarov}, {Urban},
  {Corbin}, {Wycoff}, {Bastian}, {Schwekendiek}, \& {Wicenec}}]{hog00}
{H{\o}g}, E., {et~al.} 2000, \aap, 355, L27

\bibitem[{{Hollenbach} {et~al.}(1994){Hollenbach}, {Johnstone}, {Lizano}, \&
  {Shu}}]{hollenbach94}
{Hollenbach}, D., {Johnstone}, D., {Lizano}, S., \& {Shu}, F. 1994, \apj, 428,
  654

\bibitem[{{Houck} {et~al.}(2004){Houck}, {Roellig}, {van Cleve}, {Forrest},
  {Herter}, {Lawrence}, {Matthews}, {Reitsema}, {Soifer}, {Watson}, {Weedman},
  {Huisjen}, {Troeltzsch}, {Barry}, {Bernard-Salas}, {Blacken}, {Brandl},
  {Charmandaris}, {Devost}, {Gull}, {Hall}, {Henderson}, {Higdon}, {Pirger},
  {Schoenwald}, {Sloan}, {Uchida}, {Appleton}, {Armus}, {Burgdorf},
  {Fajardo-Acosta}, {Grillmair}, {Ingalls}, {Morris}, \& {Teplitz}}]{houck04}
{Houck}, J.~R., {et~al.} 2004, \apjs, 154, 18

\bibitem[{{Hubickyj} {et~al.}(2005){Hubickyj}, {Bodenheimer}, \&
  {Lissauer}}]{hubickyj05}
{Hubickyj}, O., {Bodenheimer}, P., \& {Lissauer}, J.~J. 2005, Icarus, 179, 415

\bibitem[{{Hughes} {et~al.}(2009){Hughes}, {Andrews}, {Espaillat}, {Wilner},
  {Calvet}, {D'Alessio}, {Qi}, {Williams}, \& {Hogerheijde}}]{hughes09}
{Hughes}, A.~M., {et~al.} 2009, \apj, 698, 131

\bibitem[{{Ireland} \& {Kraus}(2008)}]{ireland08}
{Ireland}, M.~J., \& {Kraus}, A.~L. 2008, \apjl, 678, L59

\bibitem[{{Karr} {et~al.}(2010){Karr}, {Ohashi}, {Kudo}, \& {Tamura}}]{karr10}
{Karr}, J.~L., {Ohashi}, N., {Kudo}, T., \& {Tamura}, M. 2010, \aj, 139, 1015

\bibitem[{{Kenyon} {et~al.}(1998){Kenyon}, {Brown}, {Tout}, \&
  {Berlind}}]{kenyon98}
{Kenyon}, S.~J., {Brown}, D.~I., {Tout}, C.~A., \& {Berlind}, P. 1998, \aj,
  115, 2491

\bibitem[{{Kenyon} \& {Hartmann}(1995)}]{kh95}
{Kenyon}, S.~J., \& {Hartmann}, L. 1995, \apjs, 101, 117

\bibitem[{{Kim} {et~al.}(2009){Kim}, {Watson}, {Manoj}, {Furlan}, {Najita},
  {Forrest}, {Sargent}, {Espaillat}, {Calvet}, {Luhman}, {McClure}, {Green}, \&
  {Harrold}}]{kim09}
{Kim}, K.~H., {et~al.} 2009, \apj, 700, 1017

\bibitem[{{Lin} \& {Papaloizou}(1986)}]{lin86}
{Lin}, D.~N.~C., \& {Papaloizou}, J. 1986, \apj, 309, 846

\bibitem[{{Lubow} \& {D'Angelo}(2006)}]{lubow06}
{Lubow}, S.~H., \& {D'Angelo}, G. 2006, \apj, 641, 526

\bibitem[{{Luhman} {et~al.}(2010){Luhman}, {Allen}, {Espaillat}, {Hartmann}, \&
  {Calvet}}]{luhman10}
{Luhman}, K.~L., {Allen}, P.~R., {Espaillat}, C., {Hartmann}, L., \& {Calvet},
  N. 2010, \apjs, 186, 111

\bibitem[{{Mathieu} {et~al.}(1991){Mathieu}, {Adams}, \& {Latham}}]{mathieu91}
{Mathieu}, R.~D., {Adams}, F.~C., \& {Latham}, D.~W. 1991, \aj, 101, 2184

\bibitem[{{Mathis}(1990)}]{mathis90}
{Mathis}, J.~S. 1990, \araa, 28, 37

\bibitem[{{Mathis} {et~al.}(1977){Mathis}, {Rumpl}, \& {Nordsieck}}]{mathis77}
{Mathis}, J.~S., {Rumpl}, W., \& {Nordsieck}, K.~H. 1977, \apj, 217, 425

\bibitem[{{Mayer} {et~al.}(2002){Mayer}, {Quinn}, {Wadsley}, \&
  {Stadel}}]{mayer02}
{Mayer}, L., {Quinn}, T., {Wadsley}, J., \& {Stadel}, J. 2002, Science, 298,
  1756

\bibitem[{{McClure} {et~al.}(2010){McClure}, {Furlan}, {Manoj}, {Luhman},
  {Watson}, {Forrest}, {Espaillat}, {Calvet}, {DÕAlessio}, {Sargent}, {Tobin},
  \& {Chiang}}]{mcclure10}
{McClure}, M.~K., {et~al.} 2010, submitted

\bibitem[{{Monnier} \& {Millan-Gabet}(2002)}]{monnier02}
{Monnier}, J.~D., \& {Millan-Gabet}, R. 2002, \apj, 579, 694

\bibitem[{{Mulders} {et~al.}(2010){Mulders}, {Dominik}, \& {Min}}]{mulders10}
{Mulders}, G.~D., {Dominik}, C., \& {Min}, M. 2010, ArXiv e-prints
  (arXiv:1001.2146)

\bibitem[{{Muzerolle} {et~al.}(2003){Muzerolle}, {Calvet}, {Hartmann}, \&
  {D'Alessio}}]{muzerolle03}
{Muzerolle}, J., {Calvet}, N., {Hartmann}, L., \& {D'Alessio}, P. 2003, \apjl,
  597, L149

\bibitem[{{Muzerolle} {et~al.}(2004){Muzerolle}, {D'Alessio}, {Calvet}, \&
  {Hartmann}}]{muzerolle04}
{Muzerolle}, J., {D'Alessio}, P., {Calvet}, N., \& {Hartmann}, L. 2004, \apj,
  617, 406

\bibitem[{{Muzerolle} {et~al.}(2009){Muzerolle}, {Flaherty}, {Balog}, {Furlan},
  {Smith}, {Allen}, {Calvet}, {D'Alessio}, {Megeath}, {Muench}, {Rieke}, \&
  {Sherry}}]{muzerolle09}
{Muzerolle}, J., {et~al.} 2009, \apjl, 704, L15

\bibitem[{{Nagel} {et~al.}(2010){Nagel}, {D'Alessio}, {Calvet}, {Espaillat},
  {Sargent}, {Hern{\'a}ndez}, \& {Forrest}}]{nagel10}
{Nagel}, E., {D'Alessio}, P., {Calvet}, N., {Espaillat}, C., {Sargent}, B.,
  {Hern{\'a}ndez}, J., \& {Forrest}, W.~J. 2010, \apj, 708, 38

\bibitem[{{Najita} {et~al.}(2003){Najita}, {Carr}, \& {Mathieu}}]{najita03}
{Najita}, J., {Carr}, J.~S., \& {Mathieu}, R.~D. 2003, \apj, 589, 931

\bibitem[{{Norton} {et~al.}(2007){Norton}, {Wheatley}, {West}, {Haswell},
  {Street}, {Collier Cameron}, {Christian}, {Clarkson}, {Enoch}, {Gallaway},
  {Hellier}, {Horne}, {Irwin}, {Kane}, {Lister}, {Nicholas}, {Parley},
  {Pollacco}, {Ryans}, {Skillen}, \& {Wilson}}]{norton07}
{Norton}, A.~J., {et~al.} 2007, \aap, 467, 785

\bibitem[{{Nuernberger} {et~al.}(1998){Nuernberger}, {Brandner}, {Yorke}, \&
  {Zinnecker}}]{nuernberger98}
{Nuernberger}, D., {Brandner}, W., {Yorke}, H.~W., \& {Zinnecker}, H. 1998,
  \aap, 330, 549

\bibitem[{{Owen} {et~al.}(2010){Owen}, {Ercolano}, {Clarke}, \&
  {Alexander}}]{owen10}
{Owen}, J.~E., {Ercolano}, B., {Clarke}, C.~J., \& {Alexander}, R.~D. 2010,
  \mnras, 401, 1415

\bibitem[{{Paardekooper} \& {Mellema}(2006)}]{paardekooper06}
{Paardekooper}, S., \& {Mellema}, G. 2006, \aap, 459, L17

\bibitem[{{Paardekooper} \& {Mellema}(2004)}]{paardekooper04}
{Paardekooper}, S.-J., \& {Mellema}, G. 2004, \aap, 425, L9

\bibitem[{{Padgett} {et~al.}(2008){Padgett}, {Rebull}, {Stapelfeldt},
  {Chapman}, {Lai}, {Mundy}, {Evans}, {Brooke}, {Cieza}, {Spiesman},
  {Noriega-Crespo}, {McCabe}, {Allen}, {Blake}, {Harvey}, {Huard},
  {J{\o}rgensen}, {Koerner}, {Myers}, {Sargent}, {Teuben}, {van Dishoeck},
  {Wahhaj}, \& {Young}}]{padgett08}
{Padgett}, D.~L., {et~al.} 2008, \apj, 672, 1013

\bibitem[{{Pi{\'e}tu} {et~al.}(2006){Pi{\'e}tu}, {Dutrey}, {Guilloteau},
  {Chapillon}, \& {Pety}}]{pietu06}
{Pi{\'e}tu}, V., {Dutrey}, A., {Guilloteau}, S., {Chapillon}, E., \& {Pety}, J.
  2006, \aap, 460, L43

\bibitem[{{Pollack} {et~al.}(1994){Pollack}, {Hollenbach}, {Beckwith},
  {Simonelli}, {Roush}, \& {Fong}}]{pollack94}
{Pollack}, J.~B., {Hollenbach}, D., {Beckwith}, S., {Simonelli}, D.~P.,
  {Roush}, T., \& {Fong}, W. 1994, \apj, 421, 615

\bibitem[{{Pollack} {et~al.}(1996){Pollack}, {Hubickyj}, {Bodenheimer},
  {Lissauer}, {Podolak}, \& {Greenzweig}}]{pollack96}
{Pollack}, J.~B., {Hubickyj}, O., {Bodenheimer}, P., {Lissauer}, J.~J.,
  {Podolak}, M., \& {Greenzweig}, Y. 1996, Icarus, 124, 62

\bibitem[{{Pott} {et~al.}(2010){Pott}, {Perrin}, {Furlan}, {Ghez}, {Herbst}, \&
  {Metchev}}]{pott10}
{Pott}, J., {Perrin}, M.~D., {Furlan}, E., {Ghez}, A.~M., {Herbst}, T.~M., \&
  {Metchev}, S. 2010, \apj, 710, 265

\bibitem[{{Quillen} {et~al.}(2004){Quillen}, {Blackman}, {Frank}, \&
  {Varni{\`e}re}}]{quillen04}
{Quillen}, A.~C., {Blackman}, E.~G., {Frank}, A., \& {Varni{\`e}re}, P. 2004,
  \apjl, 612, L137

\bibitem[{{Rayner} {et~al.}(2009){Rayner}, {Cushing}, \& {Vacca}}]{rayner09}
{Rayner}, J.~T., {Cushing}, M.~C., \& {Vacca}, W.~D. 2009, \apjs, 185, 289

\bibitem[{{Rayner} {et~al.}(2003){Rayner}, {Toomey}, {Onaka}, {Denault},
  {Stahlberger}, {Vacca}, {Cushing}, \& {Wang}}]{rayner03}
{Rayner}, J.~T., {Toomey}, D.~W., {Onaka}, P.~M., {Denault}, A.~J.,
  {Stahlberger}, W.~E., {Vacca}, W.~D., {Cushing}, M.~C., \& {Wang}, S. 2003,
  \pasp, 115, 362

\bibitem[{{Rice} {et~al.}(2006){Rice}, {Armitage}, {Wood}, \&
  {Lodato}}]{rice06}
{Rice}, W.~K.~M., {Armitage}, P.~J., {Wood}, K., \& {Lodato}, G. 2006, \mnras,
  373, 1619

\bibitem[{{Rice} {et~al.}(2003){Rice}, {Wood}, {Armitage}, {Whitney}, \&
  {Bjorkman}}]{rice03}
{Rice}, W.~K.~M., {Wood}, K., {Armitage}, P.~J., {Whitney}, B.~A., \&
  {Bjorkman}, J.~E. 2003, \mnras, 342, 79

\bibitem[{{Rydgren} {et~al.}(1976){Rydgren}, {Strom}, \& {Strom}}]{rydgren76}
{Rydgren}, A.~E., {Strom}, S.~E., \& {Strom}, K.~M. 1976, \apjs, 30, 307

\bibitem[{{Salyk} {et~al.}(2007){Salyk}, {Blake}, {Boogert}, \&
  {Brown}}]{salyk07}
{Salyk}, C., {Blake}, G.~A., {Boogert}, A.~C.~A., \& {Brown}, J.~M. 2007,
  \apjl, 655, L105

\bibitem[{{Salyk} {et~al.}(2009){Salyk}, {Blake}, {Boogert}, \&
  {Brown}}]{salyk09}
---. 2009, \apj, 699, 330

\bibitem[{{Siess} {et~al.}(2000){Siess}, {Dufour}, \& {Forestini}}]{siess00}
{Siess}, L., {Dufour}, E., \& {Forestini}, M. 2000, \aap, 358, 593

\bibitem[{{Simon} {et~al.}(2000){Simon}, {Dutrey}, \& {Guilloteau}}]{simon00}
{Simon}, M., {Dutrey}, A., \& {Guilloteau}, S. 2000, \apj, 545, 1034

\bibitem[{{Skrutskie} {et~al.}(2006){Skrutskie}, {Cutri}, {Stiening},
  {Weinberg}, {Schneider}, {Carpenter}, {Beichman}, {Capps}, {Chester},
  {Elias}, {Huchra}, {Liebert}, {Lonsdale}, {Monet}, {Price}, {Seitzer},
  {Jarrett}, {Kirkpatrick}, {Gizis}, {Howard}, {Evans}, {Fowler}, {Fullmer},
  {Hurt}, {Light}, {Kopan}, {Marsh}, {McCallon}, {Tam}, {Van Dyk}, \&
  {Wheelock}}]{skrutskie06}
{Skrutskie}, M.~F., {et~al.} 2006, \aj, 131, 1163

\bibitem[{{Strom} {et~al.}(1989){Strom}, {Strom}, {Edwards}, {Cabrit}, \&
  {Skrutskie}}]{strom89}
{Strom}, K.~M., {Strom}, S.~E., {Edwards}, S., {Cabrit}, S., \& {Skrutskie},
  M.~F. 1989, \aj, 97, 1451

\bibitem[{{Vacca} {et~al.}(2003){Vacca}, {Cushing}, \& {Rayner}}]{vacca03}
{Vacca}, W.~D., {Cushing}, M.~C., \& {Rayner}, J.~T. 2003, \pasp, 115, 389

\bibitem[{{Varni{\`e}re} {et~al.}(2006){Varni{\`e}re}, {Blackman}, {Frank}, \&
  {Quillen}}]{varniere06}
{Varni{\`e}re}, P., {Blackman}, E.~G., {Frank}, A., \& {Quillen}, A.~C. 2006,
  \apj, 640, 1110

\bibitem[{{Ward}(1988)}]{ward88}
{Ward}, W.~R. 1988, Icarus, 73, 330

\bibitem[{{Warren}(1984)}]{warren84}
{Warren}, S.~G. 1984, \ao, 23, 1206

\bibitem[{{Weaver} \& {Jones}(1992)}]{weaver92}
{Weaver}, W.~B., \& {Jones}, G. 1992, \apjs, 78, 239

\bibitem[{{Werner} {et~al.}(2004){Werner}, {Roellig}, {Low}, {Rieke}, {Rieke},
  {Hoffmann}, {Young}, {Houck}, {Brandl}, {Fazio}, {Hora}, {Gehrz}, {Helou},
  {Soifer}, {Stauffer}, {Keene}, {Eisenhardt}, {Gallagher}, {Gautier}, {Irace},
  {Lawrence}, {Simmons}, {Van Cleve}, {Jura}, {Wright}, \&
  {Cruikshank}}]{werner04}
{Werner}, M.~W., {et~al.} 2004, \apjs, 154, 1

\bibitem[{{White} \& {Ghez}(2001)}]{white01}
{White}, R.~J., \& {Ghez}, A.~M. 2001, \apj, 556, 265

\bibitem[{{Wolf} \& {D'Angelo}(2005)}]{wolf05}
{Wolf}, S., \& {D'Angelo}, G. 2005, \apj, 619, 1114

\end{thebibliography}
\end{document}